\documentclass[compsoc,conference,a4paper,10pt,times]{IEEEtran}
\IEEEoverridecommandlockouts

\usepackage{amsmath}
\usepackage{amsthm}
\usepackage[linesnumbered,ruled,vlined]{algorithm2e}

\SetKwInOut{KwInput}{Input}
\SetKwInOut{KwOutput}{Return}

\usepackage[lambda,landau,operators,probability,sets,logic,advantage,adversary,asymptotics,notions]{cryptocode}
\usepackage{xspace}
\usepackage{bm}
\usepackage{booktabs}
\usepackage{mathtools}
\usepackage{hyperref}
\usepackage{multicol}
\usepackage{multirow}
\usepackage{graphicx}
\usepackage{xurl} 
\usepackage{siunitx}
\usepackage{pdflscape}
\usepackage[mode=buildnew]{standalone}
\usepackage{caption}
\usepackage{subcaption}
\usepackage{tikz}
\usetikzlibrary{automata,arrows,positioning,calc,fit,patterns,tikzmark,decorations.markings,shadows}
\usepackage[normalem]{ulem}
\usepackage{cleveref}
\usepackage{xargs}                      
\usepackage[colorinlistoftodos,prependcaption,textsize=tiny]{todonotes}

\theoremstyle{definition}
\newtheorem{definition}{Definition}[section]
\newtheorem{theorem}{Theorem}

\usepackage{todonotes}

\usepackage{enumitem}
\usepackage{xcolor}
\usepackage{colortbl}
\definecolor{nicegreen}{HTML}{85e085}
\definecolor{nicegreenhalf}{HTML}{D1E084}
\definecolor{plotblue}{HTML}{003366}
\definecolor{plotpink}{HTML}{FF0066}
\definecolor{plotcyan}{HTML}{00E6AC}
\definecolor{plotyellow}{HTML}{FFE066}
\usepackage{arydshln}

\newif\iffullversion
\fullversiontrue

\begin{document}
\newcommand{\etal}{\textit{et al.}}

\newcommand{\schemefont}[1]{\mathsf{#1}}
\newcommand{\procfont}[1]{\mathsf{#1}}

\newcommand*\protocolname{\textsc{Boomerang}\xspace}
\newcommand*\circled[1]{\tikz[baseline=(char.base)]{
            \node[shape=circle,draw,inner sep=2pt] (char) {#1};}}
\newcommandx{\unsure}[2][1=]{\todo[linecolor=red,backgroundcolor=red!25,bordercolor=red,#1]{#2}}
\newcommandx{\change}[2][1=]{\todo[linecolor=blue,backgroundcolor=blue!25,bordercolor=blue,#1]{#2}}
\newcommandx{\info}[2][1=]{\todo[linecolor=OliveGreen,backgroundcolor=OliveGreen!25,bordercolor=OliveGreen,#1]{#2}}
\newcommandx{\improvement}[2][1=]{\todo[linecolor=Plum,backgroundcolor=Plum!25,bordercolor=Plum,#1]{#2}}

\newcommand{\myTikzTemplate}[3]{%
    \begin{scope}[shift={(#1,#2)}, local bounding box=#3]
	\node[rect, name=rectangle#3] at (0,0) {};

	\node[circle, name=circle1#3] at (-1.5,0) {};
	\node[circle, name=circle2#3] at (-0.5,0) {};
	\node[circle, name=circle3#3] at (0.5,0) {};
	\node[circle, name=circle4#3] at (1.5,0) {};

    \end{scope}%
}

\newcommand{\ID}{\ensuremath{\mathsf{ID}}\xspace}
\newcommand{\DB}{\ensuremath{\mathsf{DB}}\xspace}
\newcommand{\mpk}{\ensuremath{\mathsf{mpk}}\xspace}
\newcommand{\msk}{\ensuremath{\mathsf{msk}}\xspace}
\newcommand{\sk}{\ensuremath{\mathsf{sk}}\xspace}
\newcommand{\pk}{\ensuremath{\mathsf{pk}}\xspace}
\newcommand{\pparams}{\ensuremath{\mathsf{pp}}\xspace}
\newcommand{\BBA}{\ensuremath{\mathsf{BBA}}\xspace}

\newcommand{\TTP}{\ensuremath{\mathsf{TTP}}\xspace}
\newcommand{\IS}{\ensuremath{\mathsf{IS}}\xspace}
\newcommand{\AC}{\ensuremath{\mathsf{AC}}\xspace}
\newcommand{\VF}{\ensuremath{\mathsf{VF}}\xspace}
\newcommand{\U}{\ensuremath{\mathsf{U}}\xspace}
\newcommand{\ATS}{\ensuremath{\mathsf{ATS}}\xspace}
\newcommand{\st}{\ensuremath{\mathsf{state}}\xspace}
\newcommand{\sttk}{\ensuremath{\mathsf{commState}}\xspace}
\newcommand{\stsig}{\ensuremath{\mathsf{sigState}}\xspace}
\newcommand{\stsp}{\ensuremath{\mathsf{spSt}}\xspace}
\newcommand{\stpl}{\ensuremath{\mathsf{plcySt}}\xspace}
\newcommand{\strw}{\ensuremath{\mathsf{rwrdSt}}\xspace}
\newcommand{\token}{\ensuremath{\mathsf{tk}}\xspace}
\newcommand{\tagn}{\ensuremath{\mathsf{tag}}\xspace}
\newcommand{\dtag}{\ensuremath{\mathsf{dtag}}\xspace}

\newcommand{\depth}{\ensuremath{\mathsf{D}}\xspace}
\newcommand{\ctree}{\ensuremath{\mathsf{CurveTree}}\xspace}

\newcommand{\n}{\sffamily\Large node n}
\newcommand{\invis}{\phantom{\sffamily\Large node n}}

\newcommand{\len}{\ensuremath{\mathsf{len}}\xspace}

\newcommand{\comm}{\ensuremath{\mathsf{Comm}}\xspace}
\newcommand{\gen}{\ensuremath{\mathsf{Gen}}\xspace}
\newcommand{\PC}{\ensuremath{\mathsf{PC}}\xspace}
\newcommand{\BPC}{\ensuremath{\mathsf{BPC}}\xspace}
\newcommand{\regu}{\ensuremath{\mathsf{RegU}}\xspace}
\newcommand{\regs}{\ensuremath{\mathsf{RegS}}\xspace}
\newcommand{\sign}{\ensuremath{\mathsf{Sign}}\xspace}
\newcommand{\verif}{\ensuremath{\mathsf{Verif}}\xspace}
\newcommand{\showgen}{\ensuremath{\mathsf{ShowGen}}\xspace}
\newcommand{\showverif}{\ensuremath{\mathsf{ShowVerif}}\xspace}
\newcommand{\setup}{\ensuremath{\mathsf{SetUp}}\xspace}
\newcommand{\eval}{\ensuremath{\mathsf{Eval}}\xspace}
\newcommand{\witcreate}{\ensuremath{\mathsf{WitCreate}}\xspace}

\newcommand{\prvr}{\ensuremath{\mathcal{P}}}
\newcommand{\vrfr}{\ensuremath{\mathcal{V}}}

\newcommand{\commit}{\schemefont{Comm}}
\newcommand{\response}{\schemefont{Resp}}
\newcommand{\challenge}{\schemefont{Chall}}

\newcommand{\Add}[1]{\textcolor{blueblind}{\uline{#1}}}
\newcommand{\Rem}[1]{\textcolor{redblind}{\uwave{#1}}}
\newcommand{\ProverAct}[1]{\textcolor{orangeblind}{\uline{#1}}}
\newcommand{\ProverCal}[1]{\textcolor{orangeblind}{#1}}
\newcommand{\VerifierAct}[1]{\textcolor{purpleblind}{\uline{#1}}}
\newcommand{\VerifierCal}[1]{\textcolor{purpleblind}{#1}}
\newcommand{\ServerAct}[1]{\textcolor{greenblind}{\uline{#1}}}
\newcommand{\ServerCal}[1]{\textcolor{greenblind}{#1}}
\newcommand{\VPShared}[1]{\textcolor{blueblind}{#1}}
\newcommand{\ClientAct}[1]{\textcolor{purpleblind}{\uline{#1}}}
\newcommand{\ClientCal}[1]{\textcolor{purpleblind}{#1}}
\newcommand{\ProverShare}[1]{\textcolor{orangeblind}{#1}}
\newcommand{\VerifierShare}[1]{\textcolor{purpleblind}{#1}}

\newcommand{\ISS}{\ensuremath{\mathtt{Issuance}}\xspace}
\newcommand{\COLL}{\ensuremath{\mathtt{Collection}}\xspace}
\newcommand{\SPN}{\ensuremath{\mathtt{Spending}}\xspace}

\newcommand{\bsa}{\schemefont{BSA}}

\newcommand{\ra}[1]{\renewcommand{\arraystretch}{#1}}
\newcommand{\curve}{\ensuremath{\mathsf{Curve}}\xspace}
\newcommand{\secp}{\ensuremath{\mathsf{secp256k1}}\xspace}
\newcommand{\secq}{\ensuremath{\mathsf{secq256k1}}\xspace}

\newtoggle{clientauth}
\usetikzlibrary{arrows.meta,calc}
\usetikzlibrary{decorations.pathreplacing,shadows}

\newcommand{\ClientAction}[2][]{
    \node[right,#1] at (\ClientX, \Y) {#2};
}
\newcommand{\VerifierAction}[2][]{
    \node[right,#1] at (\VerifierX, \Y) {#2};
}
\newcommand{\ProverAction}[2][]{
    \node[right,#1] at (\ProverX, \Y) {#2};
}
\newcommand{\ProverActionLeft}[2][]{
    \node[left,#1] at (\ProverX, \Y) {#2};
}
\newcommand{\ServerAction}[2][]{
    \node[left,#1] at (\ServerX, \Y) {#2};
}
\newcommand{\SharedAction}[2][]{
    \node[#1] at ($1/2*(\VerifierX, \Y)+1/2*(\ServerX, \Y)$) {#2}; 
}
\newcommand{\SharedActionR}[2][]{
    \node[#1] at ($1/2*(\ClientX, \Y)+1/2*(\ServerX, \Y)$) {#2}; 
}
\newcommand{\SharedVPAction}[2][]{
    \node[#1] at ($1/3*(\VerifierX, \Y)+2/3*(\ProverX, \Y)$) {#2}; 
}

\definecolor{darkgray}{rgb}{0.33, 0.33, 0.33}
\definecolor{blueblind}{HTML}{000000}
\definecolor{redblind} {HTML}{dc3220}
\definecolor{orangeblind}{HTML}{FF1800}
\definecolor{pinkblind}{HTML}{1500FF}
\definecolor{purpleblind}{HTML}{785EF0}
\definecolor{greenblind}{HTML}{1500FF}
\def\NextLineSpacing{0.45}

\newcommand{\NextLine}[1][1.0]{
  \pgfmathparse{\Y-\NextLineSpacing*#1}
  \edef\Y{\pgfmathresult}
}

\newcommand{\ClientToServer}[3][]{
  \NextLine[0.5]
  \draw[->,>=latex, #1] (\ArrowLeft,\Y) -- node[above] {#2} node[below] {#3} (\ArrowRight-0,\Y);  
  \NextLine[0.25]
}
\newcommand{\ProverToServer}[3][]{
  \NextLine[0.5]
  \draw[->,>=latex, #1] (\ArrowLeft,\Y) -- node[above] {#2} node[below] {#3} (\ArrowRight-0,\Y);  
  \NextLine[0.25]
}
\newcommand{\ServerToClient}[3][]{
  \NextLine[0.5]
  \draw[->,>=latex, #1] (\ArrowRight,\Y) -- node[above] {#2} node[below] {#3} (\ArrowLeft+0,\Y);  
  \NextLine[0.25]
}
\newcommand{\Encryption}[3][->]{
  \NextLine[0.5]
  \draw[>=latex, darkgray, #1] (\ArrowLeft,\Y) -- node[above=-0.1] {#2} node[below] {#3} (\ArrowRight,\Y);  
}

%
%
\newcommand{\StageSeparator}[1]{
  \draw[very thick,dotted,StageSeparatorColor] (\ArrowLeft,\Y-0.5*\NextLineSpacing) -- (\ArrowRight+0.15,\Y-0.5*\NextLineSpacing) node[right,anchor=west,font=\footnotesize] {stage~#1};  
}

%
%
\newcommand{\AcceptStage}[2]{
  \SharedAction{{\color{StageSeparatorColor}\textbf{accept} #2}}
  \StageSeparator{#1}
}

\newcommand{\messagerange}{.\hss.}  
\newcommand{\assignleft}{\hspace{-2pt}\leftarrow\hspace{-2pt}}

\newcommand{\circleddashed}[1]{%
  \tikz[baseline] \draw[dashed] (0,0) circle (2pt) node {#1};%
}

\title{\protocolname: A Decentralised Privacy-Preserving Verifiable Incentive Protocol}

  \author{
    \IEEEauthorblockA{
      Ralph Ankele\IEEEauthorrefmark{1},
      Sofía Celi\IEEEauthorrefmark{1},
      Ralph Giles\IEEEauthorrefmark{1},
      Hamed Haddadi\IEEEauthorrefmark{1}\IEEEauthorrefmark{2}
    }
    \IEEEauthorblockA{
      \IEEEauthorrefmark{1} Brave Software, rankele@brave.com, cherenkov@riseup.net, rgiles@brave.com, hamed@brave.com
    }
    \IEEEauthorblockA{
      \IEEEauthorrefmark{2} Imperial College London, h.haddadi@imperial.ac.uk
    }
  }

\maketitle

\begin{abstract}
In the era of data-driven economies, incentive systems and loyalty programs, have
become ubiquitous in various sectors, including advertising, retail, travel, and
financial services. While these systems offer advantages for both users and companies,
they necessitate the transfer and analysis of substantial amounts of sensitive data.
Privacy concerns have become increasingly pertinent, necessitating the development of
privacy-preserving incentive protocols. Despite the rising demand for secure and
decentralised systems, the existing landscape lacks a comprehensive solution.

In this work, we propose the \protocolname protocol, a novel decentralised
privacy-preserving incentive protocol that leverages cryptographic black box
accumulators to securely and privately store user interactions within the incentive system.
Moreover, the protocol employs zero-knowledge proofs to transparently compute rewards for users, ensuring verifiability while
preserving their privacy. To further enhance public verifiability and transparency, we utilise a
smart contract on a Layer 1 blockchain to verify these zero-knowledge proofs.
The careful combination of black box accumulators and zero-knowledge proofs makes the \protocolname protocol highly efficient.
\end{abstract}


\section{Introduction}
\label{sec:intro}

In today's data-driven world, loyalty programs and incentive systems have become key tools for companies seeking to foster enduring relationships with their users as part of customer retention strategies.
These kind of programs are getting more and more deployed around the world at a large and micro scale.
For example, Germany introduced in 2000 a payback program~\cite{PaybackGER} that currently has (as of 2024) 31 million users, and has expanded to Poland, Italy and Austria.
Switzerland introduced the Coop Supercard program~\cite{PaybackCH} and the same has happened in the UK with its Nectar loyalty program~\cite{PaybackUK}.
Typically, these systems issue users unique identifiers, often embedded in loyalty cards, which are used during transactions to accumulate or redeem points (a ``point-collection'' type of incentive system) .
The transaction and some of its details are recorded by the provider in a central account, which introduces significant privacy risks, as, more often than not, they provide the identities of their users in the clear.
However, other stored information (which includes sensitive information) can also reveal their identities.
In fact, as stated in~\cite{PoPETS:JagRup16}, \emph{Loyalty Partner}, the operator of the German payback program, received the German Big Brother Award in 2000 for violating privacy~\cite{AwardPaybackDE}.

\subsection{Our Contributions}

In this paper, we formalize the core functionality, security and privacy properties of these incentive systems, with a focus on designing a privacy-preserving solution. Building on the works of~\cite{PoPETS:JagRup16,CCS:HHNR17}, we propose \protocolname, a protocol that operates in the interactive setting and leverages a \emph{multi-use accumulative token}.
Our system allows users to accumulate and spend points privately using a token that is bound to an identifier known to both the user and the issuer (for double-spending checks).
This token is jointly issued by the user and the issuer through their secret keys.
This means that, informally, we have two parties: an user ($\U$) and an issuer.
In our design, points (positive or negative) are collected in a constant-size token in an \emph{unlinkable} manner, meaning that each point collection (or issuance) cannot be linked to another.
We use blinding and unblinding techniques to achieve this unlinkability.
When users spend their tokens, only the sum of all collected points (or a subset) and the serial number of the token are revealed to prevent double-spending.

Our protocol improves on~\cite{PoPETS:JagRup16} by providing (public) verifiability of spending transactions and avoiding the usage of pairing-based cryptography.
Moreover, in our protocol, we add the ability for other privacy-preserving attestations and to plug-in anti-fraud checks.
Additionally, Hartung \etal~\cite{CCS:HHNR17} analysed the security model of~\cite{PoPETS:JagRup16} and found a number of serious restrictions like weak security guarantees, the ability of users to \emph{pool} or trade their points between different non-expired/redeemed local tokens, and the inherit \emph{interactivity} of the system (even when claimed it was not). We incorporate fixes for those but remain in an interactive setting.
We also allow for \emph{partial} spending of the total points accumulated and for negative points to be added.
Furthermore, following the framework of~\cite{CCS:HHNR17}, we provide a security model, inspired by the indistinguishability-based security definition for a pre-payments with refunds scheme proposed in~\cite{10.1145/2699904}.
This ensures that an adversary (possibly using multiple accumulation tokens) may not be able to redeem more points than legitimately issued to them.
Our privacy definition is simulation-based and guarantees that, besides the total of accumulated points, no additional information is revealed to link transactions.

As noted, our solution is similar to that of~\cite{CCS:HHNR17}, but with some procedures changed.
We referred to the system as \emph{Accumulation Tokens System} (\ATS) and we state that it consists of 4 procedures:

First, we have an \textbf{issuance procedure} in which, each user receives $n$ multi-use accumulative tokens (henceforth, only ``tokens'') bound to their identifiers (\ID).
Note that $\U$ can receive multiple tokens and accumulate them in a global accumulator: for the purposes of our description here, we simplify it to one token.
Each token essentially consists of a (multi-)commitment $C$ and a blind signature $\sigma$ on the commitment under the issuer's secret key $\sk_{\IS}$ (the issuer is defined as $\IS$).
The commitment $C$ binds to the user's secret key ($\sk_{\U}$), a token version number ($\ID$), the balance value ($v$) (set to zero at the beginning), and some randomness ($r_1$) to be used to generate a double-spending tag in the next transaction in combination with \ID.
The associated $\pk_{\U}$ to $\sk_{\U}$, $\sigma$ and $C$ are considered public values known to $\IS$.
Note that the issuance procedure is jointly executed by $\U$ and $\IS$ so that we are assured that the received tokens are set to an initial value of 0.

The \textbf{collection procedure} consists on adding (positive or negative) points in an unlinkable manner to the received token.
One cannot simply directly send over the token to the accumulator.
Instead, the user sends a new commitment $C'$ consisting of the same ($\sk_{\U}$, $v$) as the \emph{original} token, but it resets the new token's version number $\ID'$ and randomness $r'_1$.
Then, they prove in zero-knowledge that $C'$ is indeed a new but correctly modified version of $C$, and that $C$ was honestly issued.
The double-spending tag ($\tagn$) and the version number $\ID$ will be revealed as the $\ID$ will be used to index the value $\tagn$ in an append-only database \DB.
If the accumulator ($\AC$) accepts the proofs, the homomorphic property of the commitment scheme is used to add/subtract points (the new points from a transaction) to $C'$, and the result is then blindly signed.
This new token $(C', \sigma')$ is sent to the user.

The \textbf{redemption procedure} works in a similar manner with the exception that the balance $v$ (or a subset of it) is revealed to a verifier ($\VF$), and, based on it, $\VF$ issues a reward.
For \protocolname, we add a new procedure called \textbf{(public) verifiability procedure}.
In this last procedure, the reward can be verified as calculated correctly.
This procedure is performed either with $\VF$ or publicly by using smart contracts in a L1 blockchain (a $\TTP$).

In order to realise the functionality of $\ATS$, and instead of relying on Groth-Sahai NIZK proofs and F-binding commitments (as in~\cite{CCS:HHNR17,TCC:BCKL08,IMA:IzaLibVer11}), we make use of properly defined DLOG-based NIZK proofs, Pedersen Commitments, blind signatures, and \emph{Curve Trees}~\cite{EPRINT:CamHal22}, as a type of of accumulator which does not require a trusted setup (it is completely transparent) and it is based on simple and widely used assumption: DLOG and Random Oracle Model (ROM).
By using these algorithms, we remove the usage of pairing groups (which often add a layer of extra structure) and we focus on practicality by using a construction that has short proving/verification time.
For the zero-knowledge proofs, we make use of efficient Sigma-based Proofs of Knowledge (PoK)~\cite{cryptoeprint:2023/1595} and range proofs, such as \emph{Bulletproofs}~\cite{SP:BBBPWM18}.

In a nutshell, we make the following contributions and improvements over existing works:
\begin{itemize}
	\item \textbf{Formally defined zero-knowledge proofs and blind signatures:} All our zero-knowledge proofs and signature scheme
	are properly defined, and accompanied by a formal security model and formal proofs (see~\cref{sec:buildingblocks}).
	\item \textbf{Full end-to-end implementation:} We provide a Rust implementation\footnote{\url{https://github.com/brave-experiments/Boomerang}} of the end-to-end protocol, that we open-source to allow for independent verification and widespread adoption (see~\cref{sec:implementation}).
	\item \textbf{In-depth experimental analysis:} We provide benchmarks of our end-to-end implementation and individual parts of the protocol, in a LAN/WAN setting using commodity hardware as well as on cloud-computing instances using AWS services.
	\item \textbf{Ability to perform verifiability (private and public):} \protocolname allows for verifiability of rewards either in a private or public manner. This procedure enhances the trust of users in the system while preserving their privacy (see~\cref{sec:rewards}).
	\item \textbf{Ability to plug-in anti-fraud checks:} Due to the fact that incentive systems provide financial gains for it users, they are often the target of fraud. We propose several mitigation techniques such as double-spending detection and expiration checks, to counteract
	fraudulent actors.
\end{itemize}

\iffullversion
\subsection{Background}
\label{app:back}

There are many proposals in the literature for unlinkable incentive schemes~\cite{EPRINT:CGHH10,7345331,1287337}.
However, those schemes only offer very basic features, as, for example, incentives cannot be accumulated, resulting in a system that induces high storage, and communication costs that are linear in the number of tokens.
As we have noted, from a user's point of view, incentive schemes should be privacy-preserving and should leak as little information as possible about users' behaviour.
From a issuer's point of view, the incentive scheme should prevent any fraudulent abuse of the system by preventing malicious usage in the form of double spending, for example.
These privacy-preserving properties should be maintained without impacting communication or computational costs.

A common building block for incentive systems that aims for both efficiency and privacy are Black Box Accumulators (BBA), introduced by Jager and Rupp~\cite{PoPETS:JagRup16}.
BBA-based systems offer a privacy-preserving way for point collection and point spending/redemption.
Building on this idea, the BBA+~\cite{CCS:HHNR17} scheme introduces mechanisms to prevent double spending and to accommodate negative point balances.
However, these schemes have limitations.
They rely on Groth-Sahai proofs~\cite{EC:GroSah08}, which involve cryptographic pairing groups: a structure that introduces inefficiencies in proof verification, especially at scale, where millions of users are involved.

Some incentive systems, on the contrary, aim for specialised solutions depending on the use case, and arrive to different efficiency metrics depending on the case.
For the advertisement and content recommendation case, for example, there are specific privacy-preserving advertisement delivery networks such as Adnostic~\cite{NDSS:TNBNB10} or Privad~\cite{guha2009}.
For electronic toll collection, there is~\cite{EPRINT:FHNRS18}: a system where road users accumulate debt while driving on the road, and then pay at regular intervals.
Another common use case for privacy-preserving incentive systems are anonymous payment systems.
In that context, the BBA+ scheme has been further improved to handle that use case in a system called the BBW scheme~\cite{PoPETS:HKRR20}.
The BBW system is specialised for payments on constrained devices.
Due to that, BBW drops the usage of pairing groups and replaces it with range proofs based on Bulletproofs~\cite{SP:BBBPWM18}.
It achieves unlinkability by using the Baldimtsi-Lysyanskaya blind signature~\cite{CCS:BalLys13} scheme.

Beyond incentive systems, many designs aim for general-purpose privacy-preserving solutions based on smart contract platforms~\cite{SP:KMSWP16, SP:BCGMMW20, CCS:SBGMTV19, FC:BAZB20, EPRINT:BCDF22a}.
They based their design on the usage of zero-knowledge proofs, multi-party computation and homomorphic encryption.
In many cases, private data is handled in those schemes by using Zerocash~\cite{SP:BCGGMT14}-based decentralised anonymous payment (DAP) scheme.
Other designs, on the contrary, base their solutions on trusted execution environments~\cite{8806762, 10.1145/3564625.3567995,9251926}.
While most of these schemes are general-purpose, compared to the \protocolname protocol, they are not optimised for any specific use case and therefore are often inefficient, not scalable for implementing in a large-scale system with potentially millions of users, or do not provide enough features to be used in an incentive system.
We provide comparisons with some of these systems in~\cref{sec:related}.
\fi

\paragraph{\textbf{Outline}} In~\cref{notation}, we will introduce the notations and formal framework of the cryptography used in \protocolname.
In~\cref{sec:buildingblocks}, we expand on the formal framework and provide the instantiation of the cryptography used.
Next, in~\cref{sec:boomerang-issuance} and in~\cref{sec:rewards}, we introduce the \protocolname protocol by describing its issuance, collection, spending and (public)-verifiability phases.
In~\cref{sec:analysis}, we provide a formal security and privacy analysis of \protocolname, and, in~\cref{sec:implementation}, we provide an implementation and experimental analysis of its computational and communicational costs.
Finally, in~\cref{sec:conclusion} we conclude this work.

\section{Notation}\label{notation}

We denote by $E/\mathbb{F}_p$ an elliptic curve $E$ defined over a prime field $\mathbb{F}_p$, and when its clear from the context simply write $E$ by omitting the field $\mathbb{F}_p$ for simplicity.
We denote by $p$ or $q$ the order of an elliptic curve $E$.
Unless otherwise noted, we denote group operations additively, and given a scalar $x \in \mathbb{Z}_p$ we denote with $xG$ scalar multiplication with a generator $G$.
We denote by $[n]$ the set $\{1,\ldots,n\}$.
For all intents and purposes, we consider sets to be ordered (i.e. as arrays, indexed from $1$ onwards) unless stated otherwise.
Vectors are denoted by lower-case bold letters. We use $\len(s)$ to denote the length of $s \in \bin^*$.
We write $a \leftarrow b$ to assign the value of $b$ to $a$, and $a\sample S$ to assign an uniformly sampled element from the set $S$.
$\secpar$ denotes the security parameter (arbitrary string describing general parameters: the description of a group $\mathbb{G}$ of prime order $q$ and generator $g$ of that group).
We denote negligible functions in a security parameter $\lambda$ as $\negl$.
We refer to a prover as $\prvr$ and a verifier as $\vrfr$, which are probabilistic polynomial time machines (PPTs); to a polynomial-time simulator as $\schemefont{Sim}$ and to polynomial-time algorithm extractor as $\schemefont{Ext}$.

We say that two probability ensembles $(X_n, Y_n)$, which are families of distributions over a finite set of size $n$, are:
\begin{itemize}
	\item \emph{perfectly indistinguishable}, written as $X_n \equiv Y_n$, if they are identically distributed.
	\item \emph{statistically indistinguishable}, with negligible advantage $\epsilon$, if for any computationally unbounded distinguisher $D$,
	\[\big|\Pr[D(X_n) = 1] - \Pr[D(Y_n) = 1]\big| \leq \epsilon(n)\]
	which is written as $X_n \stackrel{s}{\equiv} Y_n$.
\end{itemize}

\subsection{Homomorphic Commitment Schemes}
\label{back:comm}

A non-interactive commitment scheme allows a party to construct a commitment to a known value, so that later the commitment can be opened and the value will be revealed.
A receiver of the commitment can verify that the opened value matches the value that was committed to.
A commitment scheme must be \emph{hiding} (does not reveal the committed value) and \emph{binding} (it cannot be opened to a different value).We provide a formal definition of the scheme and its properties in~\cref{app:comm}.

\subsubsection{Pedersen Commitments}\label{pc-comm}
The Pedersen Commitment ($\PC$) scheme proposed in~\cite{C:Pedersen91} satisfies the above mentioned properties for commitment schemes.
The setup algorithm $\PC.\gen$ outputs a description of a cyclic group $\mathbb{G}$ of prime order $q$ and random generators $(G, H)$, which sets $\pparams$ to $(\mathbb{G}, q, G, H)$.
In additive notation, to commit to $m \in \mathbb{Z}_q$, a committer picks randomness $r \in \mathbb{Z}_q$ and computes $\comm(m, r) = g^mh^r$.
The scheme is perfectly hiding and computationally binding under the discrete logarithm assumption.

A variant of Pedersen Commitments allows to commit to multiple messages at once (see~\cite{ESORICS:BCCGGP15}).
This is referred to as ``multi-commitments'' or ``vector-commitment''.
Given $n$ messages $m_i \in \mathbb{Z}_q$ and generators $(g_1, \ldots, g_n)$, the commiter picks randomness $r \in \mathbb{Z}_q$, and computes $\comm(m, r) = h^r \prod_{i=1}^{n}{g_i^{m_i}}$.

\subsubsection{Blinded Pedersen Commitments}
As noted by~\cite{EPRINT:BalLys12b}, Pedersen Commitments can be further blinded.
Given $z \in \mathbb{G}$ where $z \neq 1$, $r \in \mathbb{Z}_q$, and $C = \PC.\comm((m_1, \ldots, m_n), r)$, values $(z^{\gamma}, C^{\gamma})$ can be viewed as a commitment to the same $(m_1, \ldots, m_n)$ with randomness $(r, \gamma)$.
Hence, a \emph{blinded Pedersen Commitment} ($\BPC$) is one with the following operation: $\BPC.\comm((m_1, \ldots, m_n), (r, \gamma), z) = (z^\gamma, \PC.\comm((m_1, \ldots, m_n), r)^\gamma))$.
This type of scheme preserves the binding and hiding properties.

A variant of this ``blinding'' is often referred to as \emph{rerandomisable commitments}.
Given $\delta \in \mathbb{G}$ where $\delta \neq 1$, and $C = \PC.\comm((m_1, \ldots, m_n), r)$, values $(C \cdot h^\delta)$ can be viewed as a commitment to the same $(m_1, \ldots, m_n)$ with randomness $(r, \delta)$~\cite{EPRINT:CamHal22,291060}.

\subsection{Proof of Knowledge (PoK)}\label{pok-formal}
The concept of proof of knowledge (PoK) was initially formalised by Feige, Fiat and Shamir~\cite{JC:FeiFiaSha88,C:FiaSha86}.
The construction allows for a prover $\prvr$ to convince a verifier $\vrfr$ that they know a witness $w$, which is computationally related to a common input $x$.
That is, given an NP language $\mathcal{L}$, $w$ allows a decision algorithm to determine if $x \in \mathcal{L}$ in a polynomial number of steps.
We define the relation of instance-witness pairs for a given language as a set $\mathcal{R}$.
Loosely, a proof of knowledge with \textit{knowledge error} $\kappa$ has probability $\kappa(\lambda)$ of a cheating $\prvr$ successfully convincing $\vrfr$ that $x\in \mathcal{L}$ without knowing the corresponding $w$.
The correlated language $\mathcal{L}(\mathcal{R})$ is the set of all $x$ such that there exists a $w$, where $(x,w) \in \mathcal{R}$.
We provide a formal definition of the scheme and its properties in~\cref{app:sigma}.

\subsection{Blind Signatures with Attributes}\label{subsec:blind-sigs}

The notion of \emph{blind signatures with attributes} was introduced by Baldimtsi and Lysyanskaya~\cite{EPRINT:BalLys12b}, as an extension of standard blind signatures, that allows attaching chosen user's attributes to the resulting signature via a commitment.
This means that the scheme allows signing of committed messages and verification of such signatures in an unlinkable way.
This class of signatures is commonly referred to as \emph{CL-type} signatures~\cite{SCN:CamLys02,C:CamLys04,RSA:PoiSan16,RSA:PoiSan18}, and provide both a commitment scheme and an \emph{unlinkable} blind signature scheme (relying on randomizability) that allow for signatures on committed messages, unlinkability, and security proven under interactive and non-interactive assumptions.
The signer and the user both get as input a commitment $C$ to a set of user's chosen attributes, and one can prove in zero-knowledge statements over that commitment (e.g., that it contains the correct attributes).

We describe the syntax of blind signatures with attributes, which closely follows that of~\cite{10.1145/3576915.3623184}.
A blind signature with attributes consists of three stages:
\begin{enumerate}
    \item Registration of users with signers: an user registers a commitment of their attributes and proves knowledge of the opening of them;
    \item Issuing of signatures: the user and signer interact so that the user obtains a signature on a message of their choice that is linked to the chosen attributes;
    \item Verification of the attributes and signature: this process is divided into two parts i. a signature verification algorithm, ii. two ``show'' algorithms that allow the user to reveal a subset of their attributes to a verifier, who can confirm that these attributes are linked to the signature and verify that the user knows the remaining attributes without revealing them.
\end{enumerate}
We provide a formal definition of the scheme and its properties in~\cref{app:blind}.

\subsection{2-Cycles of Curves}
\label{back:cycle}

A cycle of elliptic curves is a list of elliptic curves (in this case of length 2) defined over finite fields in which the number of points on one of the curves equals the size of the field of the next in a cyclical manner.

\begin{definition}[$m$-cycle of elliptic curves]
A $m$-cycle of elliptic curves is a list of $m$ distinct elliptic curves $E_1/\mathbb{F}_{q_{1}}, \ldots, E_m/\mathbb{F}_{q_{m}}$, where $(q_1, \ldots, q_m)$ are primes, such that the number of points on the curves satisfy:

\begin{align*} \label{cycle-points}
  &\#E_1(\mathbb{F}_{q_{1}}) = q_2, \dots,\\
  &\#E_i(\mathbb{F}_{q_{i}}) = q_{i+1}, \dots,\\
  &\#E_m(\mathbb{F}_{q_{m}}) = q_1
\end{align*}
\end{definition}
Silverman and Stange~\cite{Silverman2011} showed that arbitrary length cycles exist, and called them \emph{aliquot cycles}.
Mihailescu~\cite{mihailescu2007} analysed the case of $2$-cycles, in the context of primality proving, and called them \emph{amicable pairs}.

\subsection{Cryptographic Accumulator}\label{accumulators}

Cryptographic accumulators~\cite{EC:BenDeM93,EPRINT:DerHanSla15,cryptoeprint:2024/657} are defined as a set with a finite length of values. They provide the ability to have proofs of (non-)membership for members of said set.

\begin{definition}[Cryptographic Accumulator]
A cryptographic accumulator is a set of PPT algorithms
$\{\setup, \eval, \witcreate, \verif\}$,
defined as follows.
 \begin{itemize}
    \item $\pparams \leftarrow \setup(1^\lambda)$: takes as input the security parameter $\lambda$ and generates public parameters $\pparams$.
    \item $(acc_{\mathcal{X}}, aux) \leftarrow \eval_{\pparams}(\mathcal{X})$: computes accumulator $acc_{\mathcal{X}}$ for the input set $\mathcal{X}$, and outputs it and any optional auxiliary information $aux$.
    \item $\pi/\perp \leftarrow \witcreate_{\pparams}(\mathcal{X}, acc_{\mathcal{X}}, aux, x)$: computes a proof $\pi_{member}$ that proves that $x$ is in the set $\mathcal{X}$ if $x \in \mathcal{X}$. Otherwise, it outputs $\perp$.
    \item $1/0 \leftarrow \verif_{\pparams}(acc_{\mathcal{X}}, aux, x, \pi)$: verifies whether $x$ is in the set $\mathcal{X}$ by using the proof $\pi$. If successful, it returns 1; otherwise, 0.
\end{itemize}
\end{definition}

\section{Building Blocks}\label{sec:buildingblocks}

In this section, we describe the concrete algorithms we use in \protocolname based on the formal framework of each of them given in~\cref{notation}.

\subsection{Black-Box Accumulators as Cryptographic Accumulators}
\label{sec:bba}

Black-Box Accumulators (BBA) are a type of cryptographic accumulator, which is a data structure used to compress an underlying set.
Various schemes can be considered accumulators, starting with the foundational work of~\cite{FOCS:MicRabKil03}, that introduced the concept of \emph{zero-knowledge set}: a commitment to a hidden set that supports both membership and non-membership proofs.
Different types of accumulators exist, including those based on Groups of Unknown Order~\cite{10.1007/3-540-48285-7_24,C:CamLys02,EPRINT:BCDLRS17} or from Bilinear Pairings~\cite{RSA:LNguyen05,EPRINT:Nguyen05,EPRINT:DamTri08,AC:GOPTT16,CCS:ZBKMNS22}, and some proposals for \emph{tree of accumulators}, where every internal node is a cryptographic accumulator containing all its children~\cite{EPRINT:PapTamTri09,CCS:PamTamTri08,VerkleTree}.
This approach offers fast membership proofs but impact the accumulator's creation and update times.

When designing the \protocolname protocol, we aimed to find a highly efficient data-structure for providing zero-knowledge set-membership proofs.
We decided on Black Box Accumulators (BBAs)~\cite{PoPETS:JagRup16,CCS:HHNR17}, which allow to accumulate a finite set $\mathcal{X} =\{x_1, \dots, x_n\}$ into a \textit{succinct} value \textit{acc$_\mathcal{X}$}, as defined in~\cref{accumulators}.
Intuitively, BBAs can be seen as private counters (e.g. for incentives or points) that can only be updated by the issuer.

Following the idea of a ``tree of accumulators'', we also use \emph{Curve Trees}~\cite{EPRINT:CamHal22,291060} (\ctree) with Pedersen Commitments~\cite{C:Pedersen91} (that are \emph{rerandomizable}\cite{PKC:FleSim21} and are \emph{additively homomorphic}~\cite{AC:AbeCraFeh02}).
\ctree is a tree of cryptographic accumulators.
In the following, explain how to formalise \ctree as a cryptographic accumulator.
Note that in this subsection, we will refer to group operations additively.


\subsubsection{\ctree as a Cryptographic Accumulator}
We describe \ctree as a Cryptographic Accumulator in~\cref{app:ctree}. The formal definition follows:

\begin{definition}[$\ctree$ as a Cryptographic Accumulator] \ctree as a cryptographic accumulator is:
 \begin{itemize}
    \item $\pparams \leftarrow \setup(1^\lambda)$ takes as input the security parameter $\lambda$ and generates public parameters $\pparams$: depth factor $\depth$,
    branching factor $\ell$, a set of 2-cycle elliptic curves $(E_1/\mathbb{F}_p$, $E_2/\mathbb{F}_q$), $2n$ points in
    $E_1$ ($G_1, G_2$) and $2n$ points in $E_2$ ($H_1, H_2$).
    \item $(\ctree, \perp) \leftarrow \eval_{\pparams}(S)$ builds \ctree by using the ``Leaves construction'' and ``Nodes construction''
    steps, as defined in Definition \ref{ctree-cons}, where $\mathcal{X}$ is the set $S$ of size $n$. The result is a
    $acc_{\mathcal{X}} = \ctree$ and $aux = \perp$.
    \item $\pi_{member}/\perp \leftarrow \witcreate_{\pparams}(S, \ctree , \perp, x)$: computes a proof $\pi_{member}$ that proves that
    $x \in S$ if $x \in S$ by using~\cref{mem-proof}. Otherwise, it outputs $\perp$.
    \item $1/0 \leftarrow \verif_{\pparams}(\ctree, \perp, x, \pi_{member})$: verifies whether $x \in S$ by using the proof
    $\pi_{member}$. If successful, it returns 1; otherwise, 0.
\end{itemize}
\end{definition}

\subsection{Concrete 2-cycle Curves}
\label{sec:2-cyc}

Many instantiations of 2-cycles exists, including the Pasta cycle~\cite{pasta} or the known \secp/\secq cycle.
In \protocolname, due to their efficiency and public availability, we use the \secp/\secq cycle, but the protocol can work over any 2-cycle curves.

\subsection{Concrete Proofs of Knowledge (PoK)}\label{concrete-zkp}

In \protocolname, we use a Pedersen multi-commitment~\cite{ESORICS:BCCGGP15,EPRINT:BalLys12b} scheme~\cite{C:Pedersen91} (sometimes referred to as \emph{generalised}).
This scheme allows for a ``commit-and-prove'' technique, enabling efficient proofs of knowledge for relations where part of the witness is committed.
We require proofs of knowledge of the following relations: i. knowledge of the opening of a commitment, ii. knowledge that a value satisfies a specific arithmetic relation (such as addition, multiplication, subtraction, inner product or a combination of them), iii. knowledge that a value is a member of a set, iv. knowledge that a committed value has an specific structure.
In the following, we describe each of these proofs, noting references to existing literature where applicable.
Note that all of these proofs can be batched in a manner like~\cite{HenryRyan2014}, and we use Camenish-Stadler notation where applicable.
All of these PoK follow the framework introduced in~\cref{pok-formal} and must satisfy the properties stated there.

\subsubsection{\textbf{Proving knowledge of an opening:} $\pi_{open}(\mathbf{m})$}
\label{pi-open}

There are many techniques in the literature~\cite{EPRINT:CriLys20,PoPETS:CriLys21,PoPETS:DGSTV18} of how to create a
PoK that attests to the knowledge of the opening of a commitment.
Here we use the description given by~\cite{cryptoeprint:2023/1595}, which is based on~\cite{SP:WTSTW18,JC:Schnorr91}.
Note that in this subsection, we will refer to group operations additively.
Formally, this PoK states that, given commitment $C$ to a value $x$ with randomness $r$, $\prvr$ may convince $\vrfr$ that they possess knowledge of the value $x$ and $r$.
We expand on their description so that it works for multi-commitments, where $\mathbf{m}$ are the values of the multi-commitment:

\[
\mathcal{R}_{\text{open}} = \\
\left\{
\begin{array}{c}
(C,\pparams),\\
(\mathbf{m}, r)
\end{array}
\ \Bigg|
\begin{array}{c} \pparams = (G, (G_1,\dots,G_n,H), q),\\
    \mathbf{m} = ((m_1,\ldots, m_n),r) \\
C = r \cdot H + \sum_{k=1}^{n} m_k \cdot G_k \\
\end{array}\right\}.
\]

Concretely, given public parameters to the Pedersen multi-commitment scheme $(G_1,\dots,G_n,H)$, which are points of
an elliptic curve of order $q$ and referred to as $\pparams$, and private input $\mathbf{m} = ((m_1, \ldots, m_n),r)$, the interactive protocol works as follows.

\begin{enumerate}
    \item $\prvr$ samples $(\alpha_1, \ldots, \alpha_n), \alpha_x \sample \mathbb{Z}_q^{*}$, then computes and sends:
    \begin{align*}
        t_1 &:= \alpha_{x}\cdot H + \sum_{k\in\{n\}}{\alpha_k \cdot G_k}\end{align*}
    \item $\vrfr$ receives $t_1$, samples a challenge $c \sample \mathbb{Z}_q^{*}$, and sends it to $\prvr$.
    \item For $k\in\{n\}$, $\prvr$ computes $s_k := m_k \cdot c + \alpha_k$, and then $s_{x} := r \cdot c + \alpha_{x}$. $\prvr$ sends $((s_1, \ldots s_n), s_x)$ to $\vrfr$.
    \item $\vrfr$ accepts iff the following equation holds:
    \begin{align}\tag{i}
        c\cdot C + t_1 &\stackrel{?}{=} s_{x}H + \sum_{k\in\{n\}}{s_k \cdot G_k}
    \end{align}
\end{enumerate}
We show that $\pi_{open}(m)$ satisfies \emph{completeness}, \emph{special soundness} and
\emph{honest verifier zero-knowledge} in appendix~\ref{proof-open}.

\subsubsection{\textbf{Proving knowledge that a committed value has a specific structure:} $\pi_{issue}(\mathbf{m})$}
\label{pi-issue}

As part of the issuance procedure of \protocolname, $\prvr$ generates a commitment to the values of $\token = (\ID, 0, \sk_\U, j, r)$.
In order to guarantee that these values are correctly generated, $\prvr$ creates a proof $\pi_{issue}(\token)$ which verifies the following relation (where $\mathbf{m}$ is $\token$):

\[
  \mathcal{R}_{\text{issue}} =\\
  \left\{
  \begin{array}{c}
    (C,\pk_{\prvr},\pparams),\\
    (\mathbf{m}, r)\ \\
  \end{array}
   \Bigg|
  \begin{array}{c}
    \pparams = (G, (G_1,\dots,G_4,H),q),\\
    \mathbf{m} = (m_1,m_2,m_3,m_4,r), \\
    m_2=0,\\
    C = r \cdot H + \sum_{k=1}^{4} m_k \cdot G_k, \\
    \pk_{\prvr}=m_3 \cdot G\\
  \end{array}
  \right\}.
\]

A proof of this relation proves that: i. $\prvr$ knows the opening to the commitment $C$ which has a zero as the second entry, i.e. a counter starts at zero, ii. $\prvr$ binds any long-term identity to their commitment by including their secret key as the third entry, and proving that the commitment to $sk_{\prvr}$, and the discrete logarithm of $pk_{\prvr}$ are equal.

Given public parameters to the commitment scheme $(G_1,\dots,G_4,H)$, which are points of an elliptic curve of order $q$, the interactive
protocol works as follows. Given public input $(C,\pk_{\prvr},\pparams)$, and private input $\mathbf{m} = (m_1,m_2,m_3,m_4,r)$:

\begin{enumerate}
    \item $\prvr$ samples $\alpha_1, \alpha_3, \alpha_4, \alpha_5 \sample \mathbb{Z}_q^{*}$, then computes and sends:
    \begin{align*}
        t_1 &:= \alpha_3 \cdot G\\
        t_2 &:= \alpha_{5}\cdot H + \sum_{k\in\{1,3,4\}}{\alpha_k \cdot G_k}\end{align*}
    \item $\vrfr$ receives $(t_1,t_2)$, samples a challenge $c \sample \mathbb{Z}_q^{*}$, and sends it to $\prvr$.
    \item For $k\in\{1,3,4\}$, $\prvr$ computes: $s_k := m_k \cdot c + \alpha_k$, and then $s_{5} := r \cdot c + \alpha_{5}$. $\prvr$ sends $(s_1,s_3,s_4,s_5)$ to $\vrfr$.
    \item $\vrfr$ accepts iff the following equations hold:
    \begin{align}\tag{i}
        c\cdot \pk_{\prvr} + t_1 &\stackrel{?}{=} s_3 \cdot G\\\tag{ii}
        c\cdot C + t_2 &\stackrel{?}{=} s_{5}H + \sum_{k\in\{1,3,4\}}{s_k \cdot G_k}
    \end{align}
\end{enumerate}
We show that $\pi_{issue}(\mathbf{m})$ satisfies \emph{completeness}, \emph{special soundness} and \emph{honest verifier zero-knowledge} in appendix~\ref{proof-issuance}.

\subsubsection{\textbf{Proving that a value is the result of multiplication and addition:} $\pi_{add-mul}(m)$}
\label{pi-add-mul}

As part of the collection procedure, $\prvr$ needs to prove that certain values have been correctly generated by proving that they
are the result of both additive and multiplicative operations.
This proof ($\pi_{add-mul}(m)$) verifies the following relation:

\[\mathcal{R}_{\text{add-mul}} = \left\{
\begin{array}{c}
(C,\pparams),\\
(m, r)
\end{array} \
\Bigg|
\begin{array}{c} \pparams = (G, H, q), \\
m = ((x \cdot y) + z), \\
C = ((G \cdot m + r \cdot H) \\
\end{array}\right\}.\]

Given public parameters to the Pedersen commitment scheme $(G, H)$, which are points of an elliptic curve of order $q$, the interactive protocol works as follows. Given public input
$
(C_1 = (G \cdot x) + (H \cdot {r_1}),
C_2 = (G \cdot y) + (H \cdot {r_2}),
C_3 = (G \cdot (z)) + (H \cdot r_3),
C_4 = (G \cdot (x \cdot y)) + (H \cdot r_4),
C_5 = (G \cdot ((x \cdot y) + z)) + (H \cdot (r_4 + r_3)), \pparams)$,
and private input $((x \cdot  y)+ z, x, y, z)$:

\begin{enumerate}
    \item $\prvr$ samples $(\alpha_1, \ldots \alpha_9) \sample \mathbb{Z}_q^{*}$, then computes and sends:
    $(t_1 := \alpha_1\cdot G + \alpha_2 \cdot H), (t_2 := \alpha_3\cdot G + \alpha_4 \cdot H), (t_3 := \alpha_5\cdot G + \alpha_6 \cdot H),
      (t_4 := C_1 \cdot \alpha_3 + H \cdot \alpha_7), (t_5 := \alpha_{8}\cdot G), t_6 := \alpha_{9}\cdot H)$.
    \item $\vrfr$ receives $(t_1, \ldots, t_6)$, samples a challenge $c \sample \mathbb{Z}_q^{*}$, and sends it to $\prvr$.
    \item $\prvr$ computes: $(s_1 := c \cdot x + \alpha_1), (s_2 := c \cdot r_1 + \alpha_2), (s_3 := c \cdot y + \alpha_3), (s_4 := c \cdot r_2 + \alpha_4),(s_5 := c \cdot z + \alpha_5),(s_6 := c \cdot r_3 + \alpha_6),(s_7 := c \cdot (r_4 - r_1 \cdot y) + \alpha_7),
        (s_8 := c \cdot ((x \cdot y) + z) + \alpha_8),
        (s_9 := c \cdot (r_4 + r_3) + \alpha_{9})$
    $\prvr$ sends $(s_1, \ldots, s_5)$ to $\vrfr$.
    \item $\vrfr$ accepts iff the following equations hold:
    \begin{align}
        G \cdot {s_1} + H \cdot {s_2} &\stackrel{?}{=} t_1 + (C_1 \cdot c){}\\
        G \cdot {s_3} + H \cdot {s_4} &\stackrel{?}{=} t_2 + (C_2 \cdot c){}\\
        G \cdot {s_5} + H \cdot {s_6} &\stackrel{?}{=} t_3 + (C_3 \cdot c){}\\
        C_1 \cdot {s_3} + H \cdot {s_7} &\stackrel{?}{=} t_4 + (C_4 \cdot c){}\\
        G \cdot {s_8} + H \cdot {s_9} &\stackrel{?}{=} (t_5 + t_6) + ((C_5) \cdot c){}
    \end{align}
\end{enumerate}
We show that $\pi_{add-mul}(m)$ satisfies \emph{completeness}, \emph{special soundness} and \emph{honest verifier zero-knowledge} in appendix~\ref{proof-add-mul}.

\subsubsection{\textbf{Proving that a value is the result of subtraction:} $\pi_{sub}(m)$}
\label{pi-sub}

The aim of this proof is to attest that a committed value $m$ lies between a certain interval.
As noted in~\cite{PoPETS:HKRR20}, given an interval space $\mathbb{V} = [0, 2^{(l - 1)}]$, where $2|\mathbb{V}| < |\mathbb{Z}|$, we aim to prove
that $z = w - v \in \mathbb{V}$, which is equivalent to $w \geq v$ as long as $2|\mathbb{V}| < |\mathbb{Z}|$ or that $z$ is within the interval.
The proof convinces a verifier that a commitment $C$ opens to a number $z$ that is the provided range, without revealing $z$.
For this proof, we use range proofs and, specifically, Bulletproofs~\cite{SP:BBBPWM18}, for the relation (where $m = z$):

\[\mathcal{R}_{\text{sub}} = \left\{(C,\pparams),(m, r)\ \Bigg|
\begin{array}{c} \pparams = (G, H, q, l, \mathbb{V}), \\
C = ((G \cdot m + r \cdot H) \\
\land (z \in [0, 2^{l} - 1])) \\
\end{array}\right\}.\]

As noted, we instantiate this proof by using the range proof proposed by Bulletproofs~\cite{SP:BBBPWM18} that relies on an inner-product argument.
Bulletproofs proposes to see $z$ as $\mathbf{a} = (a_1, \ldots, a_n) \in \{0, 1\}^n$ (a vector containing the bits of $z$), so that $\langle a, 2^n \rangle = z$.
$\prvr$ commits to $\mathbf{a}$, and proves that $\langle a, 2^n \rangle = z$.
For the details of the proof, see~\cite{SP:BBBPWM18} (Section 4.1).
$\pi_{sub}(m)$ satisfies \emph{completeness}, \emph{special soundness} and \emph{honest verifier zero-knowledge} per Bulletproofs~\cite{SP:BBBPWM18}, when instantiated with it.

\subsubsection{\textbf{Proving that a value is the result of inner product:} $\pi_{reward}(\mathbf{m})$}
\label{sec:rewardproof}

The aim of this proof is to attest that a committed value $\mathbf{m}$ is the result of a inner product operation executed between two
vectors $\mathbf{(a, b)}$ and that all members of $\mathbf{m}$ do not exceed a limit $lim$.
This means that the prover knows the openings of two Pedersen commitments that satisfy a given inner product relation.
The relation, hence, is:

\[\mathcal{R}_{\text{reward}} = \left\{
\begin{array}{c}
(C,\pparams, U),\\
(a, b, \mathbf{m})
\end{array}
\ \Bigg|
\begin{array}{c} \pparams = (G, H, q, lim), \\
C = ((G \cdot a + b \cdot H + \\
(U \cdot \langle a, b \rangle)\\
\land (\mathbf{m} \in [0, lim)) \\
\end{array}\right\}.\]

We instantiate this relationship by using both the inner-product proof and range proof from Bulletproofs (Section 3 and 4, respectively),
For the details of the proof, see~\cite{SP:BBBPWM18} (Section 3 and 4).
$\pi_{reward}(m)$ satisfies \emph{completeness}, \emph{special soundness} and \emph{honest verifier zero-knowledge} per Bulletproofs~\cite{SP:BBBPWM18}, when instantiated with it.

\subsection{Concrete Blind Signatures Scheme}\label{blind-sig}

In \protocolname, we use the \emph{blind signatures with attributes} scheme introduced by Baldimtsi and Lysyanskaya~\cite{EPRINT:BalLys12b} (\textsf{ACL}), so that we don't rely on the strong RSA assumption or on pairing-based cryptography.
The construction preserves \emph{binding} as an attacker, when seeing two signatures, cannot link a signature to its issuing, and \emph{unforgeability} as an user cannot forge more signatures than what they were issued.
In the random-oracle model, the construction is unlinkable under the decisional Diffie-Hellman assumption, and unforgeable under the discrete-logarithm assumption for sequential and concurrent composition~\cite{CCS:KasLosRen23}.

In this work, we treat the generated commitments as the commitment for the signature scheme, and some underlying values of the commitment ($C$) as the message (that can be made public) for the signature.
The common inputs to an user and signer are the signer's public key for verification of signatures, and a commitment $C$ to the user's attributes $(l_1, \ldots, l_n)$.
One of those attributes will contain the user's (\U) long term identity ($\sk_\U$).
The signer's private input is its signing key, while \U's private input is the message $m$ they want signed and the opening (or the underlying values) of $C$.
\U first forms a commitment $C' = \PC.\comm((l_1, \ldots, l_n), r)$ and sends it to the signer with a $\pi_{open}(C')$.
Then, \U{} and the signer carry out the blind signature protocol relative to the combined commitment scheme.

\iffullversion
In more detail, the construction works as follows (we follow the same approach as~\cite{PoPETS:HKRR20} but corrected).
This scheme is the one from~\cite{EPRINT:BalLys12b} (ACL) and very similar to~\cite{EC:Abe01}, where signing is a three-move protocol, where knowledge of either the discrete logarithm of $\pk_\IS$ or a tagged key $z$ is proven in an OR-style manner.
The scheme assumes a common uniform random string (crs) or a random oracle that contains the values $(h_0 \ldots h_n)$, and parameters $\mathbb{G}, g, q$. $\mathcal{H}$ stands here for a hash function.

\begin{itemize}
    \item $\pparams, (\sk, \pk) \leftarrow \bsa.\mathsf{Gen(1^\lambda)}$: Samples $(x, h) \in \mathbb{Z}_p$, and generates public parameters $\pparams = \{y = g^x, z = \mathcal{H}(g, y, h)\}$ (where $z$ is the tag public key) given the security parameter $\lambda$. It sets $\sk = x$, and $\pk = y$.
    \item ($C, \pi_{{open}_{C}}) \leftarrow \bsa.\mathsf{RegU_{\pparams}(\pk, L)}$: Parses $L$ as $(l_1, \ldots, l_n)$ and computes $C = (h_0 ^ r \cdot h_1^{l_1} \cdot \cdots \cdot h_n^{l_n})$. It computes $\pi_{open}(C)$.
    \item $1/0 \leftarrow \bsa.\mathsf{RegS_{\pparams}}(\sk, C, \pi_{open}(C))$: Verifies $\pi_{open}(C)$, and outputs 1 (accept) or 0 (reject).
    \item $R \leftarrow \bsa.\mathsf{Comm_{\pparams}}(\sk, C)$: Samples $(rand, u, r_1, r_2, c) \in \mathbb{Z}_q$, and commits to those values via $z_1 = C \cdot g{^{rand}}, z_2 = z/z1, a = g^u, a_1 = g^{r_1}z_1^{c}, a_2 = h^{r_2} z_2^{c}$ in order to prove that they do not know $\log_g z_1$. It outputs and sends $\mathsf{R} = (rand, a, a_1, a_2)$.
    \item $e \leftarrow \bsa.\mathsf{Chall_{\pparams}}(\pk, R, C, m)$: Parses $\mathsf{R} = (rand, a, a_1, a_2)$, and checks that $rand \neq 0$ and that $(a, a_1, a_2) \in \mathbb{G}$. Sets $z_1 = C \cdot g^{rand}$, and generates $\gamma \in \mathbb{Z}^*_q$ and $\tau \in \mathbb{Z}_q$. Computes $\zeta = z^\gamma, \zeta_1 = z_1^\gamma, \zeta_2 = \zeta / \zeta_1$ and $\mu = z^\tau$. Generates $(t_1, t_2, t_3, t_4, t_5) \in \mathbb{Z}_q$, and calculates $\alpha = a \cdot g^{t_1} {\pk}^{t_2}, \alpha_1 = {a_1}^{\gamma} {g}^{t_3} {\zeta_1}^{t_4}$, and $\alpha_2 = {a_2}^{\gamma} h^{t_5} {\zeta_2}^{t_4}$. This step is essentially a blinding procedure and a PoK of the blinding $\mu$. Sets $\epsilon = \mathcal{H}(\zeta, \zeta_1, \alpha, \alpha_1, \alpha_4, \mu, m)$ (for a message $m$), and calculates the challenge $\mathsf{e} = \epsilon - t_2 - t_4$, which is sent.
    \item $S \leftarrow \bsa.\mathsf{Resp_{\pparams}(\sk, R, e)}$: Calculates the sub-challenges $ch = e - c \mod q$ and $r = u - ch \cdot \sk \mod q$, and sends $\mathsf{S} = (ch, c, r, r_1, r_2)$: the real and simulated answers.
    \item $(\sigma, r) \leftarrow \bsa.\mathsf{Sign_{\pparams}(\pk, R, e, S}, m)$: Parses $\mathsf{S} = (ch, c, r, r_1, r_2)$, and calculates $\rho = r + t_1 \mod q, \omega = ch + t_2 \mod q, \rho_1 = \gamma \cdot r_1 + t_3 \mod q, \rho_2 = \gamma \cdot r_2 + t_5 \mod q, \omega_1 = c + t_4$ and $\nu = \tau - \omega_1 \gamma \mod q$. If $\omega + \omega_1 = \mathcal{H}(\zeta, \zeta_1, g^{\rho} \cdot y^{\omega}, g^{\rho_1} \cdot {\zeta_1}^{\omega_1}, h^{\rho_2} \cdot {\zeta_2}^{\omega_1}, z^{v} \cdot \zeta^{\omega_1}, m)$, it computes the signature or, else, aborts. The signature is $\mathsf{\sigma} = (\zeta_1, (\zeta, \rho, \omega, \rho_1, \rho_2, v, \omega_1))$ and the opening is $\mathsf{r} = (\gamma, rand)$.
    \item $1/0 \leftarrow \bsa.\mathsf{Verif_{\pparams}(\pk, \sigma, m)}$: Parses $\mathsf{\sigma} = (\zeta_1, (\zeta, \rho, \omega, \rho_1, \rho_2, v, \omega_1))$. Outputs 1 if $\omega + \omega_1 = \mathcal{H}(\zeta, \zeta_1, g^{\rho} \cdot y^{\omega}, g^{\rho_1} \cdot \zeta_1^{\omega_1}, h^{\rho_2} \cdot {\zeta_2}^{\omega_1}, z^{v} \cdot \zeta^{\omega_1}, m)$; else, 0.
    \item $\mathsf{P} \leftarrow \bsa.\mathsf{ShowGen_{\pparams}(z, \sigma, r, L, L')}$: takes as input $z$, a signature $\mathsf{\sigma}$, the opening $\mathsf{r}$, the set of attributes $L$ and a subset of attributes $L' \subset L$. Parses $\mathsf{r}$ as $(\gamma, rnd)$ and $\mathsf{\sigma}$ as $(\zeta_1, (\zeta, \rho, \omega, \rho_1, \rho_2, v, \omega_1))$. Computes $\Gamma = g^{\gamma}$, $h' = h_i^{\gamma}$ for $i \in [n]$. Compute an equality proof $\pi_{eq}$ that shows that $\mathsf{dlog}_z \zeta = \mathsf{dlog}_g \Gamma = \mathsf{dlog}_{h_{0}} h'_0 = \cdots = \mathsf{dlog}_{h_{n}} h'_n$ (where $h_i'$ is $h_i^\gamma$). Computes the partial commitment $\zeta'_1 = \zeta_1 / h'^{\mathsf{L'}}$ and an opening proof $\pi_{open}(\zeta'_1)$, which is the rest of attributes not in $\mathsf{L'}$ with regards to $rand$ and $\mathsf{L-L'}$. Sends $\mathsf{P} = (\Gamma, h', \pi_{eq}, \pi_{open}(\mathsf{L-L'}))$.
    \item $1/0 \leftarrow \bsa.\mathsf{ShowVerif_{\pparams}(z, \sigma, L', P)}$: takes as input the tag key, the signature $\sigma$, a subset of attributes $L' \subset L$ and the proof $\mathsf{P}$. It outputs 1 (accept) or 0 (reject) if the proofs in $\mathsf{P}$ verify.
\end{itemize}
\fi

In regards to the security of the scheme,~\cite{EPRINT:BalLys12b} provided proofs for the correctness of the scheme, blindness and \emph{sequential} one-more unforgeability.
~\cite{EC:BLLOR21} proposed an efficient attack against unforgeability on the \textsf{ACL} scheme if even a logarithmic number of sessions are run \emph{concurrently}.
However, following~\cite{10.1145/3576915.3623184}, it was shown that a subtle flaw on the attack renders it inefficient as the commitment to the user's attributes is offset by $g^{rnd}$ but the DL-knowledge proof in the signature is in regards to the public key part $h$, which disallows the ability to combine sessions in a ROS-like way.
Furthermore,~\cite{10.1145/3576915.3623184} provides a proof in the algebraic group model (AGM) that the \textsf{ACL} scheme is one-more unforgeable even concurrently.
We note here that even when~\cite{PoPETS:HKRR20} claims that they have adapted the security of the scheme to be concurrently safe, this doesn't seem accurate as the scheme is already assumed safe if one does not take into account the concurrent safety of zero-knowledge proofs.
It is an open problem to analyse if the scheme is concurrently safe if one assumes the usage of a non-concurrently-safe zero-knowledge proof.

\section{\protocolname Issuance and Collection}\label{sec:boomerang-issuance}

\begin{figure*}[!ht]
    \centering
      \begin{tikzpicture}
    \edef\ProverX{-10}
    \edef\ServerX{5}

    \edef\Y{0}

    \node [rectangle,drop shadow=black,draw,fill=white,inner sep=3pt,right] at (\ProverX,\Y) {\textbf{Client}};
    \node [rectangle,drop shadow=black,draw,fill=white,inner sep=3pt,left]  at (\ServerX,\Y) {\textbf{Issuer}};

    \NextLine[1.2]
    \ServerAction{\ServerAct{static: $\pk_\IS, \sk_\IS, z$}}
    \ProverAction{\ProverAct{static: $\pk_\U, \sk_\U$}}

    \NextLine[0.7]
    \ProverAction{\ProverAct{$\st = [\token_0]$}}
    \NextLine[0.7]
    \ProverAction{\ProverAct{$\stsig = [], \sttk = []$}}

    \NextLine[0.9]
    \NextLine[0.9]
    \ProverAction{\textbf{\ISS}}
    \ServerAction{\textbf{\ISS}}

    \NextLine
    \ProverAction{\ProverCal{$\token_0 = (\ID'_0, v = 0, \sk_\U, r_1, j = 0)$}}
    \NextLine
    \ProverAction{\ProverCal{$C'_0 = \PC.\comm(\token_0)$}}
    \NextLine
    \ProverAction{\ProverCal{Generate: $\pi_{issue}(\token_0)$}}
    \NextLine
    \ProverAction{\ProverCal{$m_1 = (C'_0, \pi_{issue}(\token_0))$}}
    \NextLine
    \ProverAction{\ProverCal{$\underrightarrow{m_1}$}}
    \NextLine
    \ServerAction{\ServerCal{Verify: $\pi_{issue}(\token_0)$}}
    \NextLine
    \ServerAction{\ServerCal{$\ID''_0 \in \mathbb{Z}_p$}}
    \NextLine
    \ServerAction{\ServerCal{$C''_0 = \PC.\comm(\ID''_0, 0, 0, 0, 0)$}}
    \NextLine
    \ServerAction{\ServerCal{$C_0 = C'_0 + C''_0$}}
    \NextLine
    \ServerAction{\ServerCal{$R \leftarrow \bsa.\comm(\sk_\IS, C_0)$}}
    \NextLine
    \ServerAction{\ServerCal{$\underleftarrow{m_2 = (C_0, \ID''_0, R)}$}}
    \NextLine
    \ProverAction{\ProverCal{$C_0 = {C'}_0 + {C''}_0$}}
    \NextLine
    \ProverAction{\ProverCal{${\ID}_0 = {\ID'}_0 + {\ID''}_0$}}
    \NextLine
    \ProverAction{\ProverCal{$e \leftarrow \bsa.\challenge(\pk_\IS, R, C_0, m = \ID_0)$}}
    \NextLine
    \ProverAction{\ProverCal{$\underrightarrow{m_3 = e}$}}
    \NextLine
    \ServerAction{\ServerCal{$S \leftarrow \bsa.\response(\sk_\IS, R, e)$}}
    \NextLine
    \ServerAction{\ServerCal{$\underleftarrow{m_4 = S}$}}
    \NextLine
    \ProverAction{\ProverCal{$(\sigma_0, r_0) \leftarrow \bsa.\sign(\pk_\IS, R, e, S, m = \ID_0)$}}
    \NextLine
    \ProverAction{\ProverCal{Set $\sttk = [C_0]$ and $\stsig = [\{\sigma_0, r_0\}]$}}
    \NextLine
    \ProverAction{\ProverCal{$\token_0 = (\ID_0, 0, \sk_\U, r_1, 0)$, and rebuild $\st = [\token_0]$}}
    \NextLine
    \ProverAction{\ProverCal{Build $\ctree$ and make the root public.}}
    \NextLine
\end{tikzpicture}
    \caption{The interactive \emph{issuance procedure} of \protocolname. The client computations are illustrated in red, while the computations on the issuer side are in blue. Note that this represents the interaction with a single value.}
    \label{fig:boomeran-issuance}
\end{figure*}

Following the definition of BBA in~\cite{CCS:HHNR17} and building it as an \ATS system, a \ATS system involves five entities: a Trusted Third Party (\TTP), an Issuer (\IS), an Accumulator (\AC), a Verifier (\VF), and an User (\U).
It consists of five procedures: \emph{setup} (see~\cref{boomerang-setup}), \emph{issuance} (see~\cref{boomerang-issuance}), \emph{collection} (see~\cref{boomerang-collection}), \emph{redemption} and \emph{(public-)verification} (see~\cref{sec:rewards}.
\IS, \AC and \VF are assumed to operate under the same umbrella and share a common key-pair, indicating mutual trust.
The primary goal of the system is to generate tokens, denoted as \token, in a privacy-preserving way.
Each token is a data structure that includes the user's secret key $\sk_u$, a token version number $\ID$, a balance value $v$, a random value $r_1$ and $j$ the position of the token in a vector.
The randomness $r_1$, in combination with \ID, is used to generate double-spending tag checks.
Tokens are accumulated and spent for rewards as \U interacts with an incentive system.
A user can have many tokens as each one represents an interaction with one representative of an incentive system (each one represents the interaction with an ad, for example).
The system is privacy-preserving: \U cannot be tracked across different interactions, as their actions remain unlinked.
The only information that other parties can learn about \U is how much they spend with a specific representative of the incentive system.
By ``a specific representative of the incentive system'', we mean that the system has multiple representatives (e.g., ads in an advertising incentive system), and the user may spend tokens with one of them at a time.
Below, we explain in detail how the \emph{setup}, \emph{issuance} and \emph{collection} procedures work.
While these procedures follow the same framework outlined in~\cite{PoPETS:HKRR20}, we expand and properly detail them: each zero-knowledge proof is thoroughly constructed and defined, an accumulator (\ctree) is integrated, and anti-fraud mechanisms are addressed.
Furthermore, we highlight how anti-fraud checks can be seamlessly incorporated into each of these procedures within the system.

\subsection{Setup Procedure}\label{boomerang-setup}
In the setup procedure, \IS{} generates a long-term key pair $(\sk_{\IS}, \pk_{\IS})$, which is shared with \AC{} and \VF{}.
This long-term key pair will be used for (blind) signing purposes, which ensures that tokens are issued in a correct way without linking interactions.
Next, \U generates their own key pair $(\sk_{\U}, \pk_{\U})$, which will be used for fraud-checking procedures, as it will be used to identify \U in the system and should be bound to a physical ID (i.e. a government issued ID document such as a passport).
The public key $\pk_{\U}$ is made publicly accessible by \U.
Additionally, the user generates a vector of length $n$, denoted as \st.
Each element in this vector represents a token associated with an interaction with a representative of the incentive system, as defined by:

\[
\st = [\token_0, \token_1, \cdots \token_n]
\]

\U also creates two empty vectors, \stsig and \sttk, which will eventually store each token's blind signatures and commitments, respectively.
Contrary to the scheme presented in~\cite{CCS:HHNR17}, our system does not require a trusted setup as it is not needed for neither the accumulator nor the subsequent zero-knowledge proofs.
We also set $\depth$ (which is assumed to be constant) and $\ell \in \mathbb{N}$ for our \ctree, which depends on the length of \st.

\subsection{Issuance Procedure}\label{boomerang-issuance}
The purpose of this procedure is to jointly generate tokens that are both blindly signed and committed to.
In this procedure, \U seeks to obtain valid tokens from \IS that can be individually accumulated during interactions with the incentive system, while ensuring that the tokens cannot be used for tracking.

To begin, \U generates $n$ tokens, each initialized with a balance value $v = 0$, and places them in the \st{} vector.
Each $\token_j$ in\st{} has the following structure: a random serial number $\ID \in \mathbb{Z}_p$ (which is an additive share), the balance value $v \in \mathbb{Z}_p$ (initialised at 0), the user secret key $\sk_\U \in \mathbb{Z}_p$, unique randomness $r_1$ (which will be used for double-spending checks), and a token number $j$ that corresponds to the position of the token in \st (hence, each value can be represented as: $\token_j = (\ID, v, \sk_U, r_1, j)$).
For each $\token$ in the \st, \U generates a Pedersen commitment $C'_j$ (as stated in~\cref{pc-comm}): $n$ commitments to each $\token$ in the \st.

Formally, this means that for each $\token$ in position $j \leq n$ in the \st, it generates a Pedersen Commitment $C'_j$ to $\token_j = (\ID'_j, v_j = 0, \sk_U, {r_1}_j, j)$ by calling $C'_j = \PC.\comm(\token_j)$ for each $j \in \mathsf{len}(\st)$.
Each commitment encapsulates the token’s values in a binding and hiding way.
\U also generates zero-knowledge proofs, $\pi$, that, for each $C'_j$, prove knowledge of the following: knowledge of the long-term secret key $\sk_U$ corresponding to the public key $\pk_U$, knowledge of the underlying values of the commitment, and that the second value in each commitment is set to 0.
Several techniques can be used for this (e.g., an one-out-of-many proof~\cite{EC:GroKoh15}) and can be batched in a manner like shown in~\cite{HenryRyan2014}.
We chose to use a single zero-knowledge proof for this ($\pi_{issue}$) as shown in section~\ref{pi-issue}.
Finally, \U assembles the commitments into a vector, $\sttk' = [C'_0, C'_1, \cdots C'_n]$, where each $C'_j$ corresponds to a commitment to a \token.
\U then sends $\sttk'$, and all $n$ $\pi_{{issue}}(\token_j)_{\{j \in n\}}$.

\IS{} receives all the values, and, in turn, verifies all $n$ $\pi_{issue}(\token_j)$.
Once verified, \IS{} then generates a new random $\ID'' \in \mathbb{Z}_p$ for each $n$ in $\sttk'$, and $n$ commitments as $C''_j = \mathsf{PC.Com}(\ID''_j, 0, 0, 0, 0)$.
Then, \IS{} collects them into $\sttk'' = [C''_0, C''_1, \cdots C''_n]$, and sends all $\ID''$s and the vector $\sttk''$ to $\U$.
Next, both \IS and \U leverage the homomorphic property of Pedersen Commitments to compute $n$ new $C_j$'s, where each one is $C_j = C'_j + C''_j$.
\U also computes the final serial number $\ID_j = \ID'_j+ \ID''_j$ for each $j \in n$, which ensures that this serial number is indeed random.
For each value, then, \U and \IS use a blind signature scheme, as defined in~\cref{blind-sig} (under \IS's public key $\pk_{\IS}$), to sign each value in position $j$ in $\sttk$ ($\sttk' + \sttk''$).
Each valid signature is stored as the tuple $(\sigma, r)_j$ and added to \stsig.
Additionally, \U locally builds the \ctree by using $\sttk$ as the leaves of the tree, and makes the root of \ctree public (as in~\cref{sec:bba}).
This issuance procedure is further illustrated in~\cref{fig:boomeran-issuance} but only for a single token.

\begin{figure*}[!ht]
    \centering
      \begin{tikzpicture}
    \edef\ProverX{-10}
    \edef\ServerX{5}

    \edef\Y{0}

    \node [rectangle,drop shadow=black,draw,fill=white,inner sep=3pt,right] at (\ProverX,\Y) {\textbf{Client}};
    \node [rectangle,drop shadow=black,draw,fill=white,inner sep=3pt,left]  at (\ServerX,\Y) {\textbf{Accumulator}};

    \NextLine[1.2]
    \ServerAction{\ServerAct{static: $\pk_\IS, \sk_\IS, z, v$}}
    \ProverAction{\ProverAct{static: $\pk_\U, \sk_\U, v$}}
    \NextLine[0.7]
    \ProverAction{\ProverAct{static: $\ctree$}}
    \NextLine[0.7]
    \ProverAction{\ProverAct{$\st = [\token_0]$}}
    \NextLine[0.7]
    \ProverAction{\ProverAct{$\sttk = [C_0]$}}
    \NextLine[0.7]
    \ProverAction{\ProverAct{$\stsig = [\{\sigma_0, r_0\}]$}}

    \NextLine[0.9]
    \NextLine[0.9]
    \ProverAction{\textbf{\COLL}}
    \ServerAction{\textbf{\COLL}}

    \NextLine
    \ServerAction{\ServerCal{Generate: $\underleftarrow{m_1 = r_2 \in \mathbb{Z}_p}$}}
    \NextLine
    \ProverAction{\ProverCal{$\tagn = (\sk_U \cdot r_2) + \token_0.r_1$}}
    \NextLine
    \ProverAction{\ProverCal{Generate: $({r_1}', \ID'_0 \in \mathbb{Z}_p)$}}
    \NextLine
    \ProverAction{\ProverCal{$\token'_0 = (\ID'_0, v = \token_0.v, \sk_\U, {r_1}', \token_0.j)$}}
    \NextLine
    \ProverAction{\ProverCal{$C'_0 = \PC.\comm(\token'_0)$ and $C_{\tagn} = \PC.\comm(\tagn)$}}
    \NextLine
    \ProverAction{\ProverCal{Generate: $(\pi_{open}(C'_0), \pi_{open}(C_0), \pi_{add-mul}(C_{\tagn}), \pi_{member}(C_0))$}}
    \NextLine
    \ProverAction{\ProverCal{Generate: $P \leftarrow \bsa.\showgen(z, \sigma_0, r_0, \token_0, \token_0)$}}
    \NextLine
    \ProverAction{\ProverCal{$m_2 = (C'_0, \tagn, C_{\tagn}, \token_0.\ID, \pi_{open}(C'_{0}), \pi_{open}(C_0), \pi_{add-mul}(\tagn), \pi_{member}(\sttk[j]), \sigma_0, P)$}}
    \NextLine
    \ProverAction{\ProverCal{$\underrightarrow{m_2}$}}
    \NextLine
    \ServerAction{\ServerCal{$\bsa.\verif(\pk_\IS, \sigma_0, m = \token_0.\ID)$}}
    \NextLine
    \ServerAction{\ServerCal{$\bsa.\showverif(z, \sigma_0, \token_0, P)$}}
    \NextLine
    \ServerAction{\ServerCal{Verify: $(\pi_{open}(\token_0), \pi_{open}(\token'_0), \pi_{add-mul}(\tagn), \pi_{member}(C_0))$}}
    \NextLine
    \ServerAction{\ServerCal{$\DB[0] = \dtag = (\tagn, \token_0.\ID, r_2)$}}
    \NextLine
    \ServerAction{\ServerCal{Generate: $\ID''_0 \in \mathbb{Z}_p$}}
    \NextLine
    \ServerAction{\ServerCal{$C''_0 = \PC.\comm(\ID''_0, v, 0, 0, 0)$}}
    \NextLine
    \ServerAction{\ServerCal{$C_0 = C'_0 + C''_0$}}
    \NextLine
    \ServerAction{\ServerCal{$R \leftarrow \bsa.\comm(\sk_\IS, C_0)$}}
    \NextLine
    \ServerAction{\ServerCal{$\underleftarrow{m_3 = (C_0, \ID''_0, R)}$}}
    \NextLine
    \ProverAction{\ProverCal{$C_0 = C'_0 + C''_0$}}
    \NextLine
    \ProverAction{\ProverCal{$\ID_0 = \ID'_0 + \ID''_0$}}
    \NextLine
    \ProverAction{\ProverCal{$e \leftarrow \bsa.\challenge(\pk_\IS, R, C_0, m = \ID_0)$}}
    \NextLine
    \ProverAction{\ProverCal{$\underrightarrow{m_4 = e}$}}
    \NextLine
    \ServerAction{\ServerCal{$S \leftarrow \bsa.\response(\sk_\IS, R, e)$}}
    \NextLine
    \ServerAction{\ServerCal{$\underleftarrow{m_5 = S}$}}
    \NextLine
    \ProverAction{\ProverCal{$(\sigma_0, r_0) \leftarrow \bsa.\sign(\pk_\IS, R, e, S, m = \ID_0)$}}
    \NextLine
    \ProverAction{\ProverCal{Reset $\sttk = [C_0]$ and $\stsig = [\{\sigma_0, r_0\}]$}}
    \NextLine
    \ProverAction{\ProverCal{Rebuild $\ctree$ and make the root public.}}
    \NextLine
\end{tikzpicture}
    \caption{The collection procedure of \protocolname. The client computations are illustrated in red, while the computations on the accumulator side are in blue. Note that this represents the interaction with a single value.}
    \label{fig:boomeran-collection}
\end{figure*}

\subsection{Collection Procedure}\label{boomerang-collection}
As \U interacts with the incentive system over time (by, for example, interacting with ads), we keep track of these interactions by accumulating/collecting values (points) in $\st$ and updating them (and their commitments) accordingly.
The entity \AC accumulates these interactions.
To ensure the integrity of these updates and to prevent misuse, both \AC and \U jointly perform these updates.
For this, first, \AC chooses a random value $r_2 \in \mathbb{Z}_p$ (for clarity, we focus our explanation on updating only a single value of $\sttk$), and sends it to \U.
Upon receiving $r_2$, \U, in turn, computes a double spending tag $\tagn = (\sk_U \cdot r_2) + \token_j.r_1$, where $\token_j$ is the token being updated in $\st$.
For the $\token_j$ that \U wants to accumulate to in the $\st$, \U generates a new random value $r_1$ and a serial number $\ID' \in \mathbb{Z}_p$.
\U then generates a commitment to a new $\token_j = (\ID', v = w, \sk_U, {r_1}, j)$ by calling $C'_j = \PC.\comm(\token_j)$, where $w$ is the current balance found in $\token_j.v$ in $\st[j]$.

\U then sends the new $C'_j$, $\tagn$, $\st[j].\ID$ (the ``current'' serial number), and several proofs of knowledge: of the opening of $\pi_{open}(C'_0)$ (as in Section~\ref{pi-open}), of the opening of the ``previous'' token $\pi_{open}(C_0)$ (as in Section~\ref{pi-open}), of the correct computation of $\tagn$: $\pi_{add-mul}(\tagn)$ (as in Section~\ref{pi-add-mul}), of the correct membership of $\sttk[j]$ in \ctree{}: $\pi_{member}(\sttk[j]$) (as in Section~\ref{mem-proof}), and that the previously issued blind signature ${(\sigma, r_0)}_j$ corresponds to a previous valid signature on $\token_j$ (as in~\cref{blind-sig}).

\AC verifies all of these proofs, and, in turn, sets the double-spending tag as $\dtag = (\tagn, \sttk[j].\ID, r_2)$ and stores it locally in a database.
Next, \AC computes a new serial number $\ID'' \in \mathbb{Z}_p$, and generates a new commitment as $C'' = \PC.\comm(\ID'', v, 0, 0, 0)$, where $v$ represents the new value to accumulate.
Both \AC and \U then rely on the homomorphic property of Pedersen Commitments to compute $C = C'' + C'$.
\U will also compute ${\ID}_j = {\ID'}_j+ {\ID''}_j$.
\U and \AC then use the same blind signature scheme to sign the new commitment $C$.
For the new generated commitment $C$, \U re-builds the \ctree by replacing the commitment at the leaf level of the tree, and the same for \stsig and \sttk.
Finally, \U makes the root of the new \ctree public.
The collection procedure of \protocolname is further illustrated in Figure~\ref{fig:boomeran-collection}.

\subsection{Anti-Fraud Checks}
The design of \protocolname aims to target implementation of the protocol in large scale systems with potentially millions of users.
Therefore, and because of the nature of incentive systems providing financial gains for its users, they are prone to be misused by fraudulent actors.
Yet, we added features already at the design stage of the protocol to counter some of those threats.

\paragraph{\textbf{i. Double-Spending Detection}} Using the same token twice in a collection procedure will result in multiple transactions with the same serial number \ID{}.
To detect this, \AC{} maintains a local database of double-spending \dtag and periodically checks it for duplicate serial numbers.

During collection, \AC{} inserts each double-spending $\dtag$ into the local database.
At regular intervals, \AC{} verifies if the \ID{} inside a new received \dtag{} is already in the database ($\DB[j] = \dtag' = (tag, \ID{}, r_2$)).
If a match is found, the corresponding $\dtag'$ is retrieved from \DB and procedure aimed to identify the misbehaving user is triggered.
This procedure is designed to ensure that any user attempting to misuse the system will have their privacy protections compromised.
Given two double-spending tags $(\dtag, \dtag')$, \AC{} parses them as $(\tagn, \sttk[j].\ID, r_2)$, $(\tagn', \sttk[j].\ID', r_2')$.
\AC then checks that $(\sttk[j].\ID ==  \sttk[j].\ID') \lor (r_2 == r'_2)$, and aborts if failure.
Next, \AC calculates: $\sk_\U = ((\sk_\U \cdot r_2 + r_1) - (\sk_\U \cdot r'_2 + r_1))((r_2 - r'_2)^{-1}) \mod p$ and $\pk_U = \sk_U G$.
Finally, \AC then performs an opening proof $\pi_{open}(\sk_\U)$, which will allow to signal to a third party (together with $\pk_\U$) that the owner of $\pk_\U$ has been found misbehaving.

\paragraph{\textbf{ii. Expiration Checks}}
A key feature of the protocol is the absence of metadata associated with each token, which minimizes the risk of deanonymization and linkability.
However, this design choice makes the protocol susceptible to a ``farming'' attack, where an attacker accumulates a large number of tokens and spends or accumulates them all at once to launch a denial-of-service (DoS) attack against service providers.
This vulnerability is similar to issues observed in systems like Privacy Pass~\cite{PoPETS:DGSTV18} which have been subsequently mitigated~\cite{CCS:ChuDoHan23,EC:TCRSTW22}).
Note, however, that tokens cannot be redistributed amongst other clients as each interaction asks for a proof of knowledge of $\sk_\U$, so it is not possible to easily upgrade this attack to a Distributed-DoS.

For this attack we implement a number of possible mitigations.
First, we can set a low upper bound on the number of tokens that a single client can hold.
For instance, in an ad-incentive system, if each client can only accumulate a limited number of tokens (e.g., 100 tokens), a credible DoS attack would require considerable effort to accumulate a sufficient stockpile.
To amass 1,000,000 tokens, at least 10,000 ad interactions would be necessary.
We also recommend an expiration mechanisms to further mitigate the risk: i. periodic rotation of \IS's key so that signed tokens are implicitly tied to epochs and invalidated after each epoch, which prevents tokens from being useful beyond the validity period of their associated epoch; ii. adding an extra field to $\token_j = (\ID'_j, v_j = 0, \sk_U, {r_1}_j, j, \text{exp-date})$, so that each token is committed to an expiration date, which can be verified in zero-knowledge to ensure that it is valid only within the designated epoch.
Details of these expiration checks and their integration into the system will be explored in future work.

\section{\protocolname{} Spending and Verifiability}\label{sec:rewards}

The spending procedure is divided into two subphases: spending and (public-)verifiability.
This approach reflects scenarios where users ``spend'' points accumulated in their tokens in exchange for rewards.
In the context of digital advertising~\cite{DBLP:journals/corr/abs-2106-01940}, for instance, users earn rewards based on the number of points they've gathered, which correlates to their interactions with digital ads.
Systems like Brave Ads~\cite{bravebat} reward users with cryptocurrency in their browser wallet for engaging with advertisements.
However, users are often unaware whether the rewards they receive have been accurately computed. To address this, our system introduces the option to generate a proof that publicly verifies the correct computation of rewards.

In the first subphase, clients (\U) spend points accumulated in tokens and request rewards based on those points.
At this stage, \U proves that they are an authorised user of the system via a membership proof ($\pi_{member}$) and that the value they are spending does not exceed their balance ($\pi_{sub}$).
Optionally, they may also request a private verification to ensure that the rewards have been correctly computed.
The second subphase involves publicly verifying these rewards to prevent any miscomputation.
This verification assures clients that their rewards are accurate, addressing previous gaps in research where the reward computation relied on trusting servers to act honestly.
Below, we explain in detail how the \emph{spending} and \emph{public verifiability} procedures work, when clients interact with the verifier (\VF).
While the \emph{spending} procedure follow the same framework of~\cite{PoPETS:HKRR20}, we extend it by adding the possibility of rewards calculations and (public-)verification.

\begin{figure*}[!ht]
    \centering
      \begin{tikzpicture}
    \edef\ProverX{-10}
    \edef\ServerX{5}

    \edef\Y{0}

    \node [rectangle,drop shadow=black,draw,fill=white,inner sep=3pt,right] at (\ProverX,\Y) {\textbf{Client}};
    \node [rectangle,drop shadow=black,draw,fill=white,inner sep=3pt,left]  at (\ServerX,\Y) {\textbf{Verifier}};

    \NextLine[1.2]
    \ServerAction{\ServerAct{static: $\pk_\IS, \sk_\IS, z, v, \stpl$}}
    \ProverAction{\ProverAct{static: $\pk_\U, \sk_\U, v$}}
    \NextLine[0.7]
    \ProverAction{\ProverAct{static: $\ctree$}}
    \NextLine[0.7]
    \ProverAction{\ProverAct{$\st = [\token_0]$}}
    \NextLine[0.7]
    \ProverAction{\ProverAct{$\sttk = [C_0]$}}
    \NextLine[0.7]
    \ProverAction{\ProverAct{$\stsig = [\{\sigma_0, r_0\}]$}}

    \NextLine[0.9]
    \NextLine[0.9]
    \ProverAction{\textbf{\SPN}}
    \ServerAction{\textbf{\SPN}}

    \NextLine
    \ServerAction{\ServerCal{Generate: $\underleftarrow{m_1 = r_2 \in \mathbb{Z}_p}$}}

    \NextLine
    \ProverAction{\ProverCal{$\tagn = (\sk_U \cdot r_2) + \token_0.r_1$}}
    \NextLine
    \ProverAction{\ProverCal{Generate: $({r_1}', \ID'_0 \in \mathbb{Z}_p)$}}
    \NextLine
    \ProverAction{\ProverCal{$\token'_0 = (\ID'_0, v = \token_0.v, \sk_\U, {r_1}', \token_0.j)$}}
    \NextLine
    \ProverAction{\ProverCal{$C'_0 = \PC.\comm(\token'_0)$ and $C_{\tagn} = \PC.\comm(\tagn)$}}
    \NextLine
    \ProverAction{\ProverCal{Generate: $(\pi_{open}(C'_0), \pi_{open}(C_0), \pi_{add-mul}(C_{\tagn}), \pi_{member}(C_0), \pi_{sub}(v))$}}
    \NextLine
    \ProverAction{\ProverCal{Generate: $P \leftarrow \bsa.\showgen(z, \sigma_0, r_0, \token_0, \token_0)$}}
    \NextLine
    \ProverAction{\ProverCal{$m_2 = (C'_0, \tagn, \token_0.\ID, \pi_{open}(\token_0), \pi_{open}(\token'_0), \pi_{open}(\tagn), \pi_{member}(C_0), \pi_{sub}(v), \sigma_0, P)$}}
    \NextLine
    \ProverAction{\ProverCal{$\underrightarrow{m_2}$}}
    \NextLine

    \ServerAction{\ServerCal{$\bsa.\verif(\pk_\IS, \sigma_0, m = \token_0.\ID)$}}
    \NextLine
    \ServerAction{\ServerCal{$\bsa.\showverif(z, \sigma_0, \token_0, P)$}}
    \NextLine
    \ServerAction{\ServerCal{Verify: $(\pi_{open}(\token_0), \pi_{open}(\token'_0), \pi_{open}(\tagn), \pi_{member}(C_0),\pi_{sub}(\token(v))$}}
    \NextLine
    \ServerAction{\ServerCal{$\DB[0] = \dtag = (\tagn, \token_0.\ID, r_2)$}}
    \NextLine
    \ServerAction{\ServerCal{Generate: $\ID''_0 \in \mathbb{Z}_p$}}
    \NextLine
    \ServerAction{\ServerCal{$C''_0 = \PC.\comm(\ID''_0, v, 0, 0, 0)$}}
    \NextLine
    \ServerAction{\ServerCal{$C_0 = C'_0 - C''_0$}}
    \NextLine
    \ServerAction{\ServerCal{\fbox{Set: $\stsp[0] = v$ and calculate: $\strw = \langle \stsp, \stpl \rangle$}}}
    \NextLine
    \ServerAction{\ServerCal{\fbox{Generate: $\pi_{reward}(\strw)$}}}
    \NextLine
    \ServerAction{\ServerCal{$R \leftarrow \bsa.\comm(\sk_\IS, C_0)$}}
    \NextLine
    \ServerAction{\ServerCal{$\underleftarrow{m_3 = (C_0, \ID''_0, R, \pi_{reward}(\strw))}$}}
    \NextLine

    \NextLine
    \ProverAction{\ProverCal{\fbox{Verify: $\pi_{reward}(\strw)$}}}
    \NextLine
    \ProverAction{\ProverCal{$C_0 = {C'}_0 - {C''}_0$}}
    \NextLine
    \ProverAction{\ProverCal{${\ID}_0 = {\ID'}_0 + {\ID''}_0$}}
    \NextLine
    \ProverAction{\ProverCal{$e \leftarrow \bsa.\challenge(\pk_\IS, R, C_0, m = \ID_0)$}}
    \NextLine
    \ProverAction{\ProverCal{$\underrightarrow{m_4 = e}$}}

    \NextLine
    \ServerAction{\ServerCal{$S \leftarrow \bsa.\response(\sk_\IS, R, e)$}}
    \NextLine
    \ServerAction{\ServerCal{$\underleftarrow{m_5 = S}$}}

    \NextLine
    \ProverAction{\ProverCal{$(\sigma_0, r_0) \leftarrow \bsa.\sign(\pk_\IS, R, e, S, m = \ID_0)$}}
    \NextLine
    \ProverAction{\ProverCal{Reset $\sttk = [C_0]$ and $\stsig = [\{\sigma_0, r_0\}]$}}
    \NextLine
    \ProverAction{\ProverCal{Rebuild $\ctree$ and make the root public.}}
    \NextLine
    \NextLine
\end{tikzpicture}
    \caption{Spending and (public-)verifiability procedures of \protocolname{}. The client computations are illustrated in red, while the computations on the verifier's side are in blue. Optional steps are highlighted by a rectangular box in the diagram.}
    \label{fig:boomerang-spending}
\end{figure*}

\subsection{Spending Procedure}\label{sec:boomerang-spending}
The spending procedure works similarly as the collection procedure (in~\cref{boomerang-collection}) with the key difference that value $v$ is subtracted from the token's balance $w$ when interacting with \VF.
Additionally, another proof, $\pi_{sub}(v)$ (see Section~\ref{pi-sub}), which is a range proof, is provided to prove that $w - v \in \mathbb{V}$, privately ensuring that the value to be spent does not exceed the token's balance (note, however, that what remains private here is the token's current balance).
\VF also performs the rewards calculation and optionally provides a proof of its correct computation.
We provide details of this optional subphase/computation in~\cref{boomerang-verif}.
The spending/public verifiability procedure of \protocolname is further illustrated in Figure~\ref{fig:boomerang-spending}.

\subsection{Verification Procedure}\label{boomerang-verif}
During the spending procedure, users indicate to \VF which points they wish to spend and how many, in exchange for rewards.
They transmit these values to \VF in the form of a spending vector $\stsp$ (where each member is a $v$), after which they receive their rewards (which is a vector, as well).
Rewards are calculated as the inner-product between the spending vector and a private policy vector ($\stpl$) from \VF (one that specifies the ``value'' of each point, which is assigned by the system).
For example, if $\stsp$ at the time of reward computation is:

\[
\stsp = [0,1,3,5,0,0]
\]

and $\stpl$ is defined as:

\[
\stpl = [3,5,2,3,3,2]
\]

the reward vector ($\strw$) will be:

\begin{equation*}\label{eq:reward}
\begin{aligned}
\strw &= \langle \stsp, \stpl \rangle \\
&= 0 \times 3 + 1 \times 5 + 3 \times 2
+ 5 \times 3 + 0 \times 3 + 0 \times 2 \\
&= 26.
\end{aligned}
\end{equation*}

Optionally, \U can also obtain a proof $\pi_{reward}(\strw))$, which attests that the reward computation was performed correctly according to \VF's private policy vector (see Section~\ref{sec:rewardproof}).
As this operation can be computationally expensive, it is performed on an optional basis.
However, if at any point during the validity of the policy vector, \U wants to validate the correct computation of their reward, they can request to do so via interactions with a \TTP.
For this, \VF shares their current policy vector with \TTP, who issues a $\pi_{reward}(\strw)$ proof (see section~\ref{sec:rewardproof}).
A \TTP can be, for example, a public blockchain.
Note, however, that verifying zero-knowledge proofs as this one on a public blockchain introduces additional challenges in terms of scalability, computational effort and communication.
We have implemented a proof of concept for such a public verifiability procedure using the Solana blockchain, as outlined in Appendix~\ref{appendix:smartcontracts}.

Additional proofs might be needed from \U side: if the rewards is a fiat-currency, for example, they might want to prove that their wallet have been verified with a KYC provider, which allows to perform anti-fraud checks to attest that this is a honest user of the system.
This can be done via a ring signature, where only authorised users are part of a public ring.
In the \protocolname protocol we do not strictly require any particular KYC scheme or verification; however, we outline a possible solution for this in Appendix~\ref{appendix:identityproof}.

\paragraph{\textbf{Limitations}} Note that during spending and reward's computation, \VF{} and/or \TTP learn the value that is being spent ($v$), to which interaction it relates to (the position on $\stsp$), and the reward ($\strw$).
This leakage is inherit of the system, and we aim to mostly protect the balance that \U holds and make interactions unlinkable.
It remains as future work to create a system that protects the spent value, to which interaction it relates to, and the reward value.

\section{Security Analysis}\label{sec:analysis}

There are several goals that we aim for:

\begin{itemize}
  \item \textbf{Identity Binding:} Tokens can only be issued to and used (accumulated or spent) for their legitimate user as they are binded to their identities.
  \item \textbf{Balance Binding:} \U cannot claim more value than they have collected up to that point.
  \item \textbf{Double-spending checks:} Users presenting expired/spent versions of their tokens can be identified of this misbehaviour.
  \item \textbf{User privacy:} Users are assured that their operations remain unlinkable and will not be used for tracking, unless they misbehave. Collusion between \IS, \AC, and \VF should be unable to link the issuance process to the accumulation or spending one.
  \item \textbf{False-accusation protection:} Users are assured that they cannot be accused of double-spending if they have not done so.
  \item \textbf{Verifiability:} Users are assured that their reward was correctly computed.
\end{itemize}

\subsection{Formal Security}

Our security model for \protocolname and its procedures (\emph{issuance}, \emph{collection}, and \emph{spending} and \emph{public verifiability}) is based on the security models presented in~\cite{10.1145/2699904, CCS:HHNR17,PoPETS:HKRR20}.
As so, our security model is game-based and our privacy model is simulation-based.

We define the advantage of an algorithm $\mathcal{A}$ in \protocolname for a security game \emph{Game} by:
\begin{equation*}
\text{Adv}_{\mathcal{A}, \protocolname}^{\text{game}}(\lambda) = \Pr[\mathcal{A}~\text{succeeds}] - \Pr[\mathcal{A}~\text{fails}]
\end{equation*}
and we say that the protocol is secure with respect to the security game,
iff $\text{Adv}_{\mathcal{A}, \protocolname}^{\text{game}}(\lambda) \le \text{negl}(\lambda)$ is negligible
for a security parameter $\lambda$.

\subsubsection{Security Model of \protocolname}
We define our above mentioned properties
by means of experiment in presence of a probabilistic polynomial time adversary $\mathcal{A}$. The adversary may control a set
of malicious colluding dishonest users and may also eavesdrop on honest users of the incentive system. Additionally, the adversary
may concurrently interact with an honest user \U, issuer \IS, accumulator \AC, and verifier \VF an arbitrary number of times, only bounded by the
polynomial runtime. Furthermore, the adversary may manipulate the messages sent during the protocol flow of malicious users (note, the adversary only acts as a passive adversary to honest users).

The adversary $\mathcal{A}$ can query the following oracles, in addition to executing the normal protocol flow of honest users:

\begin{description}
\label{oracles}
\item[MalIssue(pk$_{\U}$):] this oracle initiates the \emph{issuance procedure} with an honest user $\U$ and the issuer $\IS$. To ensure that we only have a single token
issued per user public key pk$_{\U}$, we required that pk$_{\U}$ has not been used in any call to either MalIssue(pk$_{\U}$) or the honest
issuance procedure.
\item[MalCollect(v):] this oracle initiates the \emph{collection procedure} to accumulate points in the incentive scheme between the
adversary $\mathcal{A}$ and the accumulator $\AC$.
\item[MalSpendVerify(v, w):] this oracle initiates the \emph{spending and public verifiability procedure} between the adversary $\mathcal{A}$
and the verifier $\VF$.
\end{description}

Note, there is additionally a \emph{setup procedure} to set up keys and other system parameters that we assume are already been
determined and we do not include in the description of the oracles due to brevity.

\begin{theorem}[Security and Privacy of \protocolname]
\protocolname is secure and privacy with respect to an ideal incentive system, when assuming the following:
\begin{itemize}
	\item the hardness of the discrete-log (DLOG) problem used by the underlying building blocks: for the zero-knowledge proofs (see~\cref{concrete-zkp}), \ctree (see~\cref{sec:bba}), commitment scheme (see~\cref{pc-comm}), and blind signature scheme (see~\cref{blind-sig}),
	\item special-soundness and zero-knowledge of the zero-knowledge proofs used (see~\cref{concrete-zkp}),
	\item unforgeability and unlinkability of the blind signature scheme used (see~\cref{blind-sig}),
	\item the following procedures are owner-binding, balance-bindig and ensure double-spending detection, when an adversary is given access to the oracles in~\cref{oracles}, given their:
        \begin{itemize}
	   \item underlying security of \emph{issuance procedure}~\cite{10.1145/2699904},
	   \item underlying security of \emph{collection procedure}~\cite{10.1145/2699904},
	    \item underlying security of \emph{spending and public verifiability procedure}, which is the same as the \emph{collection} one~\cite{10.1145/2699904}, augmented with the special-soundness and zero-knowledge of Bulletproofs~\cite{SP:BBBPWM18},
	\end{itemize}
\end{itemize}
\end{theorem}

A proof for the above theorem follows a standard hybrid argument, where the procedures of \protocolname are replaced with the ideal functionality in a step-by-step fashion, showcasing that the same result can be achieved by the real-world \protocolname.
Finally, when all the ideal procedures are in place, the rest of the security proof follows from the properties of the underlying building blocks.
As a consequence, the security of \protocolname is given by the cryptographic primitives chosen and used.
For the specific experiments of the oracle experiments following~\cite{10.1145/2699904}, see~\cref{app:args}.
Furthermore, we provide some extra security considerations in~\cref{sec-cons}.

\section{Experimental Analysis}\label{sec:implementation}

\paragraph{\textbf{Implementation}} We implemented a prototype of \protocolname in Rust in around 25k lines of code\footnote{\url{https://github.com/brave-experiments/Boomerang}}.
Our implementation consists of: an implementation of the \textsf{ACL} signature scheme~\cite{EPRINT:BalLys12b}, a zero-knowledge proof module for our proofs as an extension of the \textsf{CDLS} library~\cite{cdls}, a reimplementation of the \textsf{Bulletproofs} library~\cite{bulletproofs} extended to allow for 2-cycle curves from the \textsf{arworks} library~\cite{arkworks}, some e2e example functionality, and a core implementation of our protocol.
We think that our implementation of the different functionality as separate modules can be of independent interest and a contribution on its own, because, as far as we are aware, our implementation contains primitives that are not available elsewhere.
We also provide a full listing of the changes made to third party libraries alongside our prototype.

\paragraph{\textbf{Experimental Parameters}} For our experiments, we use \secp and \secq as our 2-cycle of curves, which provide a security level of 128 bits.
We experimented only with the interaction with one incentive system that is collected to and spent/verified (the length of \st is, hence, 1).
This, in turn, sets the $\depth$ and $\ell$ from \ctree to 1 (we also provide numbers for \ctree with $n = 2^{20}$, $\ell = 32$ and $\depth = 2$  in~\cref{table:indiv-bandwidth}).

\paragraph{\textbf{Computational Setup}} For estimating the runtime of \protocolname, we use two AWS EC2 instances: (i) Intel(R) Xeon(R) CPU E5-2686 v4 @ 2.30GHz, 32GiB of memory (referred in AWS EC2 as ``t2.2xlarge'') for server-side operations, and (ii) Intel(R) Xeon(R) CPU E5-2686 v4 @ 2.30GHz, 16GiB of memory and 4vCPU (referred in AWS EC2 as ``t2.xlarge'').
Both machines are fairly small, which highlights the practicality of our solution.
The results are found on~\cref{table:aws-bandwidth}.
In addition, we provide benchmarks using a Macbook M3 with 24GB of memory (\cref{table:bandwidth}), to highlight efficiency of operations on commodity hardware.
The experiments use single-thread execution and results are taken as the average of 100 runs.

\begin{table}[t]
\ra{1.2}
\centering
  \resizebox{\columnwidth}{!}{%
\begin{tabular}{l r r r r r}
    \toprule
    {\(\curve\)} & {\(\textbf{M1}\)} & {\(\textbf{M2}\)} & {\(\textbf{M3}\)} & {\(\textbf{M4}\)} & {\(\textbf{M5}\)} \\
    \midrule
    \multicolumn{6}{c}{\textbf{Issuance}} \\
    \midrule
    \textbf{\secp/\secq} & 525 & 455 & 399 & 160 & {-} \\
    \multicolumn{6}{c}{\textbf{Total = 1539}} \\
    \midrule
    \multicolumn{6}{c}{\textbf{Collection}} \\
    \midrule
    \textbf{\secp/\secq} & 32 & 2463 & 487 & 399 & 160 \\
    \multicolumn{6}{c}{\textbf{Total = 3541}} \\
    \midrule
    \multicolumn{6}{c}{\textbf{Spending}} \\
    \midrule
    \textbf{\secp/\secq} & 32 & 7578 & 487 & 399 & 160 \\
    \multicolumn{6}{c}{\textbf{Total = 8656}} \\
    \midrule
    \multicolumn{6}{c}{\textbf{Spending-Verify}} \\
    \midrule
    \textbf{\secp/\secq} & 32 & 7578 & 10046 & 399 & 160 \\
    \multicolumn{6}{c}{\textbf{Total = 18215}} \\
    \bottomrule
\end{tabular}
}
\caption{Bandwidth costs (\(\SI{}{\byte}\)) for \protocolname. Note that they are the same for the two curves.}
\label{table:bandwidth-t}
\end{table}

\begin{table}[t]
\centering
  \resizebox{\columnwidth}{!}{%
  \begin{tabular}{l c r r r r r}
      \toprule
      {\(\curve\)} & {\(\textbf{M1}\)} & {\textbf{M2}} & {\textbf{M3}} & {\textbf{M4}} & {\textbf{M5}} &{\textbf{Populate-state}}\\
      \midrule
      \multicolumn{7}{c}{\textbf{Issuance}}\\
      \midrule
  \textbf{\secp} & 1.6170 & 2.4436 & 1.9297 & 4.9476e-5 & \text{-} & 1.2547 \\
  \textbf{\secq} & 1.7198 & 2.6453 & 2.0897 & 4.999e-5 & \text{-} & 1.3688 \\
  \midrule
  \multicolumn{7}{c}{\textbf{Collection}}\\
  \midrule
  \textbf{\secp} & 0.0031248 & 6.9421 & 7.9561 & 1.9376 & 4.8443e-5 & 1.2533 \\
  \textbf{\secq} & 0.003133  & 7.4639 & 8.5464 & 2.0989 & 4.8545e-5 & 1.3358 \\
  \midrule
  \multicolumn{7}{c}{\textbf{Spending}}\\
  \midrule
  \textbf{\secp} & 0.0031645 & 71.120 & 22.832 & 2.7481 & 6.0415e-5 & 1.4491 \\
  \textbf{\secq} & 0.0031239 & 75.345 & 12.560 & 2.346 & 5.10544e-5 & 1.1774\\
  \bottomrule
  \multicolumn{7}{c}{\textbf{Spending-Verify}}\\
  \midrule
  \textbf{\secp} & 0.0031675 & 70.180 & 82.196 & 7.7455 & 5.0415e-5 & 1.2491 \\
  \textbf{\secq} & 0.0031239 & 75.536 & 88.155 & 8.3250 & 5.1098e-5 & 1.3774\\
  \bottomrule
  \end{tabular}
  }
\caption{Performance (\(\SI{}{\milli\second}\)) on a single client with a ``t2.xlarge'' machine and a server with a ``t2.2xlarge'' machine, for \protocolname.}
\label{table:aws-bandwidth}
\end{table}

In terms of other performance metrics, we use the current AWS financial cost structure for running a server in the ``t2.2xlarge''~\cite{aws-cost}.
Therefore, the CPU per-hour cost is estimated as \$0.1856/4 = \$0.0464 (since this machine has 4 vCPUs, and we run single-threaded), the download cost is \$0.09 per GB\footnote{As of September 2024 for the first 10TB}, and the upload cost is zero.
The results of our calculations are shown in~\cref{table:financial-t}.

\begin{table}[t]
  \ra{1.2}
\centering
  \resizebox{\columnwidth}{!}{%
  \begin{tabular}{c | r r r}
      \toprule
      {\textbf{Procedure}} & {\textbf{Runtime ($\SI{}{\milli\second}$)}} & {\textbf{Download ($\SI{}{\byte}$)}} & {\textbf{Total Cost (USD)}} \\
      \midrule
  \textbf{Issuance} & 2.4436 & 615 & 0.000051 \\
  \textbf{Collection} & 7.9593 & 679 & 0.00005700 \\
  \textbf{Spending} & 22.8352 & 679 & 0.00005719 \\
  \textbf{Spending-Verify} & 82.1992 & 10238 & 0.000859 \\
  \bottomrule
  \end{tabular}}
\caption{Financial costs (\$USD) for \protocolname. Note that they are for \secp.}
\label{table:financial-t}
\end{table}

We evaluated the performance of our prototype from both the bandwidth (\cref{table:bandwidth-t}) and runtime (\cref{table:aws-bandwidth}) perspective.
As noted, our core protocol consists of four procedures (one is optional), each one with at most 5 messages sent (only Issuance -~\cref{sec:boomerang-issuance}- consists of 4 messages).
We refer to each message sent as $\schemefont{MX}$ for $\schemefont{X} = \{1, \ldots, 5\}$.
The process by which \U populates its local state (by executing the last step of signing, re-building \ctree, re-populating \sttk and \stsig) is referred to as ``$\schemefont{Populate-state}$''.

\begin{table}[t]
  \centering
  \resizebox{\columnwidth}{!}{%
    \begin{tabular}{c|c|c}
        \toprule
        \multirow{2}{*}{\shortstack{\textbf{Number of users}}} & \multicolumn{2}{c}{\shortstack{\textbf{System Performance}}} \\
        ~ & $\mathsf{localhost}$ & $\mathsf{AWS EC2}$ \\
        \midrule
        1 & 2.12 & 3.25 \\
        10 & 3.50 & 6.12 \\
        100 & 9.34 & 20.34 \\
        1000 & 83.00 & 332.00 \\
         \bottomrule
    \end{tabular}
    }
  \caption{Performance of the end-to-end implementation for multiple users (in \(\SI{}{\second}\)). It is measured by running multiple client instances and a single-threaded server instance on both localhost and an AWS EC2 ``t2.2xlarge'' instance.}
   \label{tab:end2endbenchmarks}
\end{table}

\subsection{Analysis and Financial Cost Estimates}
As noted, we analyse the end-to-end performance of \protocolname in two scenarios: i. a single client, ii. multiple clients interacting with the system.
We also provide the ability to publicly verify the computation of rewards (as noted in~\cref{boomerang-verif}) in a centralised backend server or in a decentralised smart contract on the Solana blockchain (where we mimicked a validator by using a test server with an 2.8Ghz AMD EPYC 7402P 24-Core Processor and 256GB memory).
We also provide individual numbers of each primitive that is part of the protocol in~\cref{table:indiv-bandwidth}.

As stated, the runtime results of the usage of \protocolname by a single client are reflected on~\cref{table:aws-bandwidth}.
We note that the spending (\cref{sec:boomerang-spending}) and spending-verify (\cref{boomerang-verif}) procedures are the most costly, but we argue that the periodicity by which they are executed is less than the issuance or collection procedures: a user might ask only monthly, for example, for their rewards.
Bandwidth costs are highlighted on~\cref{table:bandwidth-t}, and we see the same reasoning as when detailing the runtime costs.

We implemented an end-to-end version of the \protocolname protocol as well, as we aimed to analyse our protocol's implementation in a multi-user setting, running several client instances and a single-threaded server instance.
We run the protocol over TLS, and the results are found on~\cref{tab:end2endbenchmarks}.
We present both the results of running the server instance on a ``localhost'' machine (Apple M3, 24GB of memory), and those when using an AWS EC2 instance (i.e. in a WAN environment:``t2.2xlarge'').
For better performance and scalability, the server instance should be multi-threaded, either implemented as serverless functions on an appropriate backend, or using containerised applications such as Kubernetes.
However, we leave this as future work.
As seen in the table, the costs scale linearly with the number of clients, which is expected.

The more costly parts of the protocol are both the spending and spending-verify procedures, as they use more costly zero-knowledge proofs (as seen in~\cref{table:indiv-bandwidth}).
However, as we have noted, some of the costly operations are optional, and do not need to be executed in a regular basis.
Furthermore, they can be executed by a \TTP, such as a L1 blockchain.
We discuss the choice of L1 blockchain and costs of using it in~\cref{appendix:smartcontracts}.

Finally, we want to provide some financial cost estimates.
We calculate the server-side costs for running an incentive system based on the \protocolname protocol, for two possible deployment scenarios: i) when the rewards proof verification is done on an AWS EC2 backend (as seen in~\cref{table:financial-t}), and ii) when the rewards verification is done on-chain on a L1 blockchain within a smart contract (see Table~\ref{tab:costs} in~\cref{appendix:smartcontracts}).
In both scenarios, the rest of the server-side computations in the Issuance, Collection and Spending procedures are done on an AWS EC2 instance.
We provide a comparison with related works from the experimental perspective in~\cref{sec:related}.

\iffullversion
\section{Discussion}

\subsection{\textbf{Related work}}\label{sec:related}

As noted throughout, \protocolname builds on top of the work of the seminal work of ``Black-Box Accumulators''~\cite{PoPETS:JagRup16} (BBA),
``Improved Black-Box Accumulators''~\cite{CCS:HHNR17} (BBA+) and ``Black-Box Wallets''~\cite{PoPETS:HKRR20} (BBW).
They represent the most efficient accumulative incentive systems to date.
Systems like Privacy Pass~\cite{PoPETS:DGSTV18, challenge-bypass} can also be used as incentive systems.
In~\cref{tab:comparison}, we compare \protocolname with BBA+ and BBW, by using their reported experimental results in their paper (Section 7.2 of ~\cite{PoPETS:HKRR20}).
However, note that the measurements reported for BBA+ and BBW are the result of running those protocols on a modern smartphone featuring a Snapdragon 845, and with an implementation written in C++, while the ones from \protocolname come from~\cref{table:aws-bandwidth} and the Rust implementation.
Their implementation also uses 16-bit range proofs and Curve25519, which might be faster than our \secp curve.
As neither the BBA+ nor the BBW protocols have an open source implementation, we were unable to reproduce their results in the same machine that \protocolname used.
Note that we do not compare directly with BBA as BBA+ is the improved version of it so we prioritise the latter.
We also \emph{only} report measurements for \protocolname for ``Spending-Verify'' as it is the only protocol that provides this procedure.

\begin{table*}[h!]
    \centering
    \begin{tabular}{l|rrrrrrr}
        \toprule
        {\textbf{Procedure}} &\multicolumn{3}{r}{\textbf{Execution time ($\SI{}{\milli\second}$)}} &\multicolumn{3}{r}{\textbf{Data transmitted ($\SI{}{\byte}$)}} & {\textbf{\# of rounds}}\\
        \midrule
        {} &{\textbf{BBA+}} & {\textbf{BBW}} &{\textbf{\protocolname}} &{\textbf{BBA+}} & {\textbf{BBW}} &{\textbf{\protocolname}} & \\
        \midrule
        \textbf{Issuance} & 172 & \cellcolor{nicegreenhalf}72 & \cellcolor{nicegreen}\(\num{7.25}\) & \cellcolor{nicegreenhalf}1024 & \cellcolor{nicegreen}1005 & 1539 & 2\\
        \textbf{Collection} & 546 & \cellcolor{nicegreenhalf}107 & \cellcolor{nicegreen}16.84 & 4208 & \cellcolor{nicegreen}1745 & \cellcolor{nicegreenhalf}3541 & $2 \& 1/2$ \\
        \textbf{Spending} & 1312 & \cellcolor{nicegreenhalf}304 & \cellcolor{nicegreen}98.15 & 14368 & \cellcolor{nicegreen}3921 & \cellcolor{nicegreenhalf}8656 & $2 \& 1/2$ \\
        \textbf{Spending-Verify} & {-} & {-} & 161.37 & {-} & {-} & 18215 & $2 \& 1/2$ \\
        \bottomrule
    \end{tabular}
    \caption{Comparison with BBA+ and BBW. Green and light green indicate the most and second-most optimal cases.}
    \label{tab:comparison}
\end{table*}

As we see in~\cref{tab:comparison}, \protocolname performs the best from an runtime perspective, and it is the second best in the majority of the subprotocols in regards to data transmitted across the network.
For the ``Issuance'' and ``Collection'' procedures, for example, \protocolname is an order of magnitude quicker, and it is a half-order of magnitude quicker for ``Spending''.

\subsection{Security Considerations}
\label{sec-cons}

\subsubsection{Key consistency}

A potential attack on \protocolname's unlinkability and \U's privacy is when the server (either acting as \IS, \AC or \VF) uses different long-term secret keys ($\IS_{\sk}$, for example) for different clients.
A server that does this can then correlate issuance with collection or redemption, for example.
To prevent this kind attack, we can ensure that the server commits to the key it uses each time and that it publicly advertises the key (for example, by placing it in append-only logs or in the Tor consensus).
A PoK (of the form~\cite{PoPETS:DGSTV18}) can be used to assure users that the secret part of the publicly advertised key is the one used in the protocol's execution.

\subsubsection{Key-rotation policies}

To prevent compromise, it is important that servers (either acting as \IS, \AC or \VF) and clients implement a rotation policy of the long-term key material.
However, this can render previously signed tokens invalid, so an implementation of \protocolname should be careful on its rotation policy.

\subsubsection{Leakage} As discussed in previous sections, \protocolname leaks the amount that \U spend, the rewards they get and the incentive they are getting a reward for (that they watched, for example, an ad $x$ times and got rewarded $y$ value).
This leakage is inherit of the system, and we are aiming to protect the balance that users have.
It remains as future work to derive a solution that diminishes or eliminates this leakage.

\else
\fi

\section{Conclusion}\label{sec:conclusion}

In this work, we proposed a novel privacy-preserving incentive protocol (\protocolname) that enables decentralized, publicly verifiable rewards redemption without requiring a trusted setup by using pairing-free elliptic curves for zero-knowledge proofs. Our choice of an L1 blockchain, supporting parallel smart contract execution, ensures scalability for millions of users. We demonstrate the feasibility of our protocol through an end-to-end proof of concept, with performance metrics indicating its suitability for widespread use. Future work includes optimizing privacy by shifting reward verification to the client, implementing elliptic-curve agnostic smart contracts, and exploring privacy-preserving analytics for advertisers.


\iffullversion
\section*{Acknowledgement}
We would like to thank Shai Levin for his help with the various aspects of the zero-knowledge proofs, and Alex Davidson for useful conversations.
\else
\fi

\bibliographystyle{IEEEtranS}
\bibliography{local.bib,cryptobib/crypto.bib,cryptobib/abbrev0.bib}

\appendices

\section{Formal definitions}

\subsection{Homomorphic Commitment Schemes}
\label{app:comm}

Formally, following the definition of~\cite{EC:GroKoh15}, a non-interactive commitment scheme is a pair of probabilistic polynomial time algorithms $(\gen, \comm)$.
The setup algorithm $\pparams \leftarrow \gen(1^\lambda)$ generates public parameters $\pparams$.
The parameters $\pparams$ determine a message space $\mathcal{M}_{\pparams}$, a randomness space $\mathcal{R}_{\pparams}$ and a commitment space $\mathcal{C}_{\pparams}$.
The commitment algorithm $\comm$ combined with $\pparams$ determines a function $\comm_{\pparams} : \mathcal{M}_{\pparams} \times \mathcal{R}_{\pparams} \rightarrow \mathcal{C}_{\pparams}$.
Given a message $m \in \mathcal{M}_{\pparams}$, the sender picks uniformly at random $r \sample \mathcal{R}_{\pparams}$ and computes the commitment $C = \comm(m, r)$.

\begin{definition}[Commitment Scheme]
A non-interactive commitment scheme $\Gamma$ consists of probabilistic polynomial time algorithms $C = \{\gen, \comm\}$ defined as follows:
\begin{itemize}
    \item $\pparams \leftarrow \Gamma.\gen\mathsf{(1^\lambda)}$: generates public parameters $\pparams$ given the security parameter $\lambda$.
    \item $C \leftarrow \Gamma.\comm_{\pparams}(m, r)$: computes a commitment $C$ for message $m$ and randomness $r$. As $r$ can be generated as part of this algorithm, it will be sometimes omitted.
\end{itemize}
\end{definition}

For simplicity we write $\comm = \comm_{\pparams}$. It must provide the following properties:

\begin{definition}[Hiding]
A non-interactive commitment scheme $\Gamma$ is \emph{hiding} if a commitment does not reveal the value it was committed to.
For all probabilistic polynomial time stateful adversaries $\adv$:
  \[
    \condprob{\substack{\pparams \leftarrow \Gamma.\gen\mathsf{(1^\lambda)}; \hfill \\ (m_0, m_1) \leftarrow \adv(\pparams); \hfill \\ b \leftarrow \{0, 1\}; \hfill \\ C \leftarrow \Gamma.\comm(m_b)}}{\adv(C) = b} \approx \frac{1}{2},
  \]
  where \adv{} outputs $m_0, m_1 \in \mathcal{M}_{\pparams}$. If the probability is exactly $\frac{1}{2}$, we say $\Gamma$ is \emph{perfectly hiding}.
\end{definition}

\begin{definition}[Biding]
A non-interactive commitment scheme $\Gamma$ is \emph{biding} if a commitment cannot be opened to a different value it has been committed to.
For all probabilistic polynomial time stateful adversaries $\adv$:
  \[
    \condprob{
    \substack{
    \pparams \leftarrow \Gamma.\gen\mathsf{(1^\lambda)}; \hfill \\
     \substack{(m_0, r_0, \\
     \hspace{0.2cm} m_1, r_1)}
     \leftarrow \adv(\pparams)
     }
     }
    {\substack{m_0 \neq m_1 \land \hfill \\ \Gamma.\comm(m_0, r_0) =  \Gamma.\comm(m_1, r_1)}
    } \approx 0,
  \]
  where \adv~outputs $m_0, m_1 \in \mathcal{M}_{\pparams}$ and $r_0, r_1 \in \mathcal{R}_{\pparams}$.
  If the probability is exactly $0$, we say $\Gamma$ is \emph{perfectly biding}.
\end{definition}

In this paper, we are interested in commitment schemes that are homomorphic, which means that the commitment space is also a group (written multiplicatively with $\cdot$).
For all well-formed $\pparams$ and $(m_0, m_1) \in \mathcal{M}_{\pparams}$ and $(r_0, r_1) \in \mathcal{R}_{\pparams}$, we have that $\comm(m_0, r_0) \cdot \comm(m_1, r_1) = \comm(m_0 + m_1, r_0 + r_1)$.

\subsubsection{Sigma Protocols}
\label{app:sigma}

A sigma protocol $\Sigma$ is an interactive protocol between two parties $\prvr$ and $\vrfr$.
We provide a brief description below, and further details can be found at~\cite{Hazay2010,dam-sigma,prop-sigma}.

Given a common input, where both $\prvr$ and $\vrfr$ have $x$, and $\prvr$ has a value $w$ (the witness) such that $(x, w) \in \mathcal{R}$, a $\Sigma$-protocol for a relation $\mathcal{R}$ works as follows:

\begin{description}
  \item[\textbf{Round 1 (Commit)}:] $\mathcal{P}$ sends a message $\commit$ to $\vrfr$ .
  \item[\textbf{Round 2 (Challenge)}:] On receiving $\commit$, $\vrfr$ sends a random public coin $t$-bit string $\challenge$.
  \item[\textbf{Round 3 (Response)}:] $\mathcal{P}$ receives the challenge and sends a reply $\response$ to $\vrfr$.
  \item[\textbf{Verification:}] $\mathcal{V}$ receives $\response$ and decides to accept or reject based solely on seen data. $\mathcal{V}$ outputs 1 if it accepts; otherwise, 0.
\end{description}

We assume that $\prvr$'s only advantage over $\vrfr$ is that they know the private $w$.
We stress that parties use independent randomness for generating their messages in every execution. They must satisfy the following properties:

\begin{description}
\item[Perfect Completeness.]  If $(x, w) \in \mathcal{R}$, and $\mathcal{P}$ and $\mathcal{V}$ follow the protocol honestly, on input $x$ and $\prvr$'s private input $w$, then $\mathcal{V}$ accepts with probability 1.
\item[Special Honest Verifier Zero-Knowledge (HVZK).] When given $x$ and a challenge, $\schemefont{Sim}$ outputs a valid transcript that is (perfectly) indistinguishable from a real transcript (with the same probability distribution as those between a honest $\prvr$ and a honest $\vrfr$ on common input $x$). Note, that the challenge being randomly sampled implies that $\vrfr$ behaves honestly.
\item[Special Soundness.] When given any $x$ and any pair of accepting distinct transcripts $[(\commit, \challenge, \response)$, $(\commit, \challenge', \response')]$ for $x$, where $\challenge \neq \challenge'$, $\schemefont{Ext}$ outputs $w$ so that $(x,w) \in \mathcal{R}$. This property~\cite{pub-21438} is restricted to the case where \emph{two} colliding transcripts are necessary and sufficient for extracting $w$. However, this property can be relaxed to \emph{$n$-special soundness}~\cite{EPRINT:GroKoh14,ESORICS:BCCGGP15,EPRINT:Wikstrom18,EPRINT:Wikstrom21,EPRINT:AABOR21b}, where $\schemefont{Ext}$ needs $n>2$ colliding transcripts to extract $w$.
\item[$n$-Special Soundness~\cite{TCC:AttFehKlo22}.] Given $x$ and $n$ valid distinct transcripts $[(\commit_i, \challenge_i, \response_i)_{i \in [n]}]$ where $\commit_i = \commit_j$ (with a common first message), $\challenge_i \neq \challenge_j$ for all $1 \leq i < j \leq n$, $\schemefont{Ext}$ outputs $w$ such that $(x, w) \in \mathcal{R}$.
An $n$-special sound $\Sigma$-protocol with a $t$-bit challenge space has knowledge error $\frac{n-1}{2^t}$~\cite[Eqn. 1]{TCC:AttFehKlo22}. It is known that $n$-special-soundness for $\Sigma$-protocols implies knowledge soundness and thus renders PoKs.
\end{description}

The Fiat-Shamir transform~\cite{C:FiaSha86} allows for converting an interactive $\Sigma$-protocol into a non-interactive one by replacing $\vrfr$'s challenges with the output of a random oracle query on the protocol transcript~\cite{CCS:BelRog93,JC:PoiSte00}.
In the random oracle model (ROM), the transformation preserves knowledge soundness and the protocol is perfect Zero-Knowledge (ZK) if the underlying $\Sigma$-protocol is perfect HVZK.
Note, however, that even though the proof is now non-interactive, the extractor for HVZK works by rewinding, which might not be ideal for some protocols~\cite{10.1007/BFb0054113}.
One can use an \emph{online extractor}~\cite{C:Fischlin05}, to circumvent this, specially, if concurrent executions of a protocol are expected.

As previously noted, PoK arrive to the HVZK property, where the verifier is assumed to be honest.
However, in practical applications, this may be insufficient as a malicious verifier could issue dishonest challenges.
It is easy to convert a HVZK argument into a full zero-knowledge one (following~\cite{EC:Damgard00}) that is secure against arbitrary verifiers in the common reference string model using standard techniques.
The conversion is very efficient and only incurs a small additive overhead.
In the following, we denote zero-knowledge proofs with $\pi_{ops}(\text{val})$, where $ops$ indicates the operation being proven and $val$ represents the value related to the proof.
When expressing an NP relation ($\mathcal{R}(x,w)$) we use Camenisch-Stadler notation~\cite{Camenisch1997ProofSF}.

\subsection{Blind Signature with Attributes}
\label{app:blind}

\begin{definition}[Blind Signature with Attributes]
A blind signature with attributes ($\bsa$) is a set of algorithms
$\{\gen, \regu, \regs, \comm, \challenge, \response, \sign, \verif,
\\
\showgen, \showverif\}$,
defined as follows.

\begin{itemize}
    \item $\pparams, (\sk, \pk) \leftarrow \bsa.\mathsf{Gen(1^\lambda)}$: generates public parameters $\pparams$ and a key pair $(\sk, \pk)$ given $\lambda$.
    \item ($C, \pi_{open}(C)) \leftarrow \bsa.\regu_{\pparams}(\pk, \textbf{l})$: takes as input $\pk$ and a vector of attributes $\textbf{l}$. It computes a commitment $C$ to $\textbf{l}$ with randomness $r$, and an opening proof $\pi_{open}(C)$.
    \item $ 1/0 \leftarrow \bsa.\regs_{\pparams}(\sk, C, \pi_{open}(C))$: takes as input $\sk$, a commitment $C$ and $\pi_{open}(C)$. It verifies $\pi_{open}(C)$, and outputs 1 (accept) or 0 (reject).
    \item $R \leftarrow \bsa.\comm_{\pparams}(\sk, C)$: takes as input $\sk$, and a commitment $C$. Outputs a commitment $R$.
    \item $e \leftarrow \bsa.\challenge_{\pparams}(\pk, R, C, m)$: takes as input $\pk$, commitments $(C, R)$, and a message $m$. Outputs challenge $e$.
    \item $S \leftarrow \bsa.\response_{\pparams}(\sk, R, e)$: takes as input $\sk$, $R$, and a challenge $e$. Outputs a response $S$.
    \item $(\sigma, r) \leftarrow \bsa.\sign_{\pparams}(\pk, R, e, S, m)$: takes as input $\pk$, $R$, challenge $e$, response $S$ and the message. Outputs a signature $\sigma$ and an opening $r$.
    \item $1/0 \leftarrow \bsa.\verif_{\pparams}(\pk, \sigma, m)$: takes as input $\pk$, a signature $\sigma$ and the message $m$. It outputs 1 (accept) or 0 (reject).
    \item $\pi_{show} \leftarrow \bsa.\showgen_{\pparams}(\pk, \sigma, r, \textbf{l}, \textbf{l'})$: takes as input $\pk$, a signature $\sigma$, an opening $r$, the vector of attributes $\textbf{l}$ and a sub-vector of attributes $\textbf{l'} \subset \textbf{l}$. It outputs a proof $\pi_{show}(\textbf{l - l'})$.
    \item $1/0 \leftarrow \bsa.\showverif_{\pparams}(\pk, \sigma, \textbf{l'}, \pi_{show})$: takes as input $\pk$, a signature $\sigma$, an opening $r$, a subset of attributes $\textbf{l'} \subset \textbf{l}$ and a proof $\pi_{show}$. It outputs 1 (accept) or 0 (reject).
\end{itemize}
\end{definition}

\subsection{\ctree as a Cryptographic Accumulator}
\label{app:ctree}

\ctree is a construction introduced in~\cite{EPRINT:CamHal22,291060} as a cryptographic accumulator that allows for efficient set-membership proofs.
The construction starts by first building a shallow Merkle Tree (with a small depth)\footnote{Note that the construction is not a ``Merkle Tree'' per se as it does not uses hashes, but it is very similar.} where the leaves and internal nodes are commitments to elliptic curve points.
The leaves and nodes are constructed as follows:
A \emph{2-cycle of elliptic curves} (see~\cref{back:cycle}) consists of the following curves ($E_{1}, E_{2}$) and two prime fields $(\mathbb{F}_p, \mathbb{F}_q)$ for two primes $(p, q)$, so that $p = |E_{1}/\mathbb{F}_p|$ and $q = |E_{2}/\mathbb{F}_q|$, and the scalar field of $E_1$ equals the base field of $E_2$ and vice versa.
Note, that a point $A$ (with coordinates $(a_x, a_y)$) in $E_1/\mathbb{F}_p$ can be represented as a pair of scalars in $E_2$ so that $a_x \cdot B + a_y \cdot C \in E_2$ for points $(B, C) \in E_2$ (this concept generalises to any number of $l$ points).
This can be, in turn, interpreted as (non-hiding) Pedersen Commitments in $E_2$ of points from $E_1$.
With this in mind, the leaf level of the Merkle Tree has points in $E_1$, the subsequent level generates the node Pedersen Commitments in $E_2$ from the leaf level (of $n$ points, which will determine the branching factor $\ell$) for each subtree, and so on until the root of the tree.
Note, that the interpreted Pedersen Commitments are \emph{non-hiding}, but they can be extended to be so, by sampling an additional point $H$ per curve, which is then used to randomise a commitment by computing $C' = C + r \cdot H$, as noted in~\cref{back:comm}.
\ctree uses algebraic operations and linear verification assuming the representation of group elements as \emph{pairs of scalars for
a (distinct) group}, which provides nuances to~\cite{TCC:CFGG22}.

\begin{definition}\label{ctree-cons}
A \ctree is defined by a depth factor $\depth$, a branching factor $\ell$, a set of 2-cycle elliptic curves $(E_1/\mathbb{F}_p$, $E_2/\mathbb{F}_q$), $2n$ points in $E_1$ ($G_1, G_2$) and $2n$ points in $E_2$ ($H_1, H_2$).
\ctree is constructed recursively over $\depth$ as follows (starting from $\depth = 0$):

\begin{itemize}
    \item \textbf{Leaves construction:} A $(\depth = 0, \ell, E_1, E_2)$---\ctree.
    Here, leaf nodes (as a set $S$ which size is $n$) are instantiated as points in $E_1$, and are labelled as $C_{0, i}$ for $i \leq n$.
    The set of leaf nodes $S$ will be partitioned in subsets of $\ell$ size which will represent a \ctree.
    \item \textbf{Nodes construction:} A $(\depth + 1, \ell, E_1, E_2)$---\ctree.
    Here, internal parent node(s) as level \depth are a list of their $\ell$ children $(\depth-1, \ell, E_2, E_1)$---\ctree.
    Each child is labelled as $C_{D-1, i}$ for $i \leq \ell$.
    The label of the internal parent node itself is defined as $C_{\depth, i} = ({C_{\depth-1, i}}_{x} \cdot G_1 + {C_{D-1, i}}_{y} \cdot G_1)$ for all $i$.
    Note that at each level $k \leq \depth$, the curve $E_j$ is switched for $j \in \{1, 2\}$.
    This means that each node is a Pedersen Commitment of its $\ell$ children, where the root is a commitment to all children.
\end{itemize}

\end{definition}

Let us consider a simple example of a $(\depth = 3, \ell = 2, E_1, E_2)$---\ctree to help our understanding, as illustrated in Figure~\ref{fig:curvetree}.
The circles in Figure~\ref{fig:curvetree} correspond to elliptic curve points in $E_j$ for $j \in \{1, 2\}$, where the elliptic curves are a 2-cycle.
The rounded rectangles are Pedersen Commitments to those elliptic curve points.
When constructing a \ctree for a set $S$ (the set of all leaf points where $|S| = n$), we can do as follows:
we start with $S$ as the leaves of the tree, we partition $S$ in smaller subsets of size $\ell$ and create a \ctree for each subset.
We recursively build the \ctree at each level until the root, where the parent nodes are Pedersen Commitments of the children (of each subset).

\begin{figure}
    \centering
    \scalebox{0.9}{
\begin{tikzpicture}
\begin{scope}[font=\small,draw,rectangle, every node/.style={fill=black!10,shape=rectangle,draw,minimum size=1.3cm,pattern=dots},
  level 1/.style={sibling distance=2cm},
  level 2/.style={sibling distance=4cm},
  level 3/.style={sibling distance=10cm,scale=1.5},
  emph/.style={edge from parent/.style={->,red,thick,draw,sloped,pattern=none}},
  ]
\node[draw=none] (TREE) {\textbf{R:} $C_{\depth, 0}$}
child { node[draw=none] {$C_{\depth-1, 0}$}
child { node[draw=none] {$C_{\depth-2, 0}$}
child [emph] { node[draw=none] {$PC_0$} edge from parent node[fill=none,draw=none,below,yshift=12pt] {PtoPC}
}
}
child {node[draw=none] {$C_{\depth-2, 1}$}
child [emph] { node[draw=none] {$PC_2$} edge from parent node[fill=none,draw=none,below,yshift=12pt] {PtoPC}
}
}
}
child {node[draw=none] {$C_{\depth-1, 1}$}
child {node[draw=none] (D2) {$C_{\depth-2, 2}$ }
child [emph] { node[draw=none] (PC1) {$PC_1$} edge from parent node[fill=none,draw=none,below,yshift=12pt] {PtoPC}
}
}
child {node[draw=none] {$C_{\depth-2, 3}$}
child [emph] { node[draw=none] (D3) {$PC_3$} edge from parent node[fill=none,draw=none,below,yshift=12pt] {PtoPC}
}
}
};
\end{scope}
\begin{scope}[->,font=\small,draw,circle, every node/.style={fill=white!10,shape=circle,draw},
  edge from parent/.style={black,thick,draw},
  level 1/.style={sibling distance=2cm},
  level 2/.style={sibling distance=4cm}]
\node (TREE) {\textbf{R:} $C_{\depth, 0}$}
child { node {$C_{\depth-1, 0}$}
child {node {$C_{\depth-2, 0}$}}
child {node {$C_{\depth-2, 1}$}}
}
child {node (D1) {$C_{\depth-1, 1}$}
child {node {$C_{\depth-2, 2}$}}
child {node (D2) {$C_{\depth-2, 3}$}}
};
\node[right =of TREE,draw,densely dashed,draw=none] (e1) {\textbf{$\in E_1$}};
\node[right =of D1,draw,densely dashed,xshift=-20pt,draw=none] (e2) {$\in E_2$};
\node[right =of D2,draw,densely dashed,xshift=-20pt,draw=none] (e1) {$\in E_1$};
\end{scope}
\end{tikzpicture}
}
\caption{Example of a $(\depth=3, \ell=2, E_1, E_2)$---\ctree. The circles indicate elliptic curve points, while the rounded rectangles indicate Pedersen commitments to those elliptic curve points: the first row just takes the ``original points'' and creates Pedersen commitments from them (by using the PtoPC operation), which returns other points.
Then, at each level we take the points in $E_1$ and use their coordinates $(x, y)$ to transform them to points in $E_2$ and vice versa at each level.}
\label{fig:curvetree}
\end{figure}

\subsubsection{Membership Proof of \ctree}\label{mem-proof}
Extending the fact that a \ctree can be made \emph{hiding} as each node in the tree can be randomized by using rerandomisable commitments, it is possible to prove membership of a value on a \ctree in zero-knowledge way by descending the tree one level at a time starting from the root.
The procedure starts by opening a commitment to $(\depth, \ell, E_1, E_2)$---$\ctree$.
Next, one picks one of its children (in zero-knowledge), rerandomize the child and output the resulting hiding commitment to the $(\depth-1, \ell, E_2, E_1)$---$\ctree$, and so recursively.
For this, one intuitively needs a opening-proof of the parent node of the chosen child, and then an opening proof of the child that has been rerandomised (with $\delta$, as $C^{*} = C + (H \cdot \delta) \in E_2$, in additive notation).
Note, that this strategy works for as many levels but will require proofs for each level of the \ctree.
A more efficient strategy exists: a single proof can be created at once and for multiple layers that work in the same group, which, when using 2-cycle curves, reduces the algorithm to two proofs: one for the parents at odd layers and one for parents at even layers.
This membership proof ($\pi_{member}$) is more formally defined in~\cite{EPRINT:CamHal22,291060}.

\section{Security of PoK schemes}\label{zkp-proofs}

\subsection{Security of $\pi_{issue}(\mathbf{m})$}\label{proof-issuance}

The PoK $\pi_{issue}(\mathbf{m})$ works as detailed in Section~\ref{pi-issue}.
Below, we prove its properties.

\subsubsection{Completeness} Completeness holds since the verification equations are always satisfied for a valid witness, and hence the verifier accepts with probability 1.

\subsubsection{Special Soundness} Given two colliding transcripts for the same instance $(C,\pk_{\prvr},\pparams)$, we construct a valid witness for the given relation. Denote the colliding transcripts as: \[(t_1,t_2,c,\{s_k\}_{k\in\{1,3,4,5\}}) \text{ and } (t_1,t_2,c',\{s'_k\}_{k\in\{1,3,4,5\}}).\] By taking the difference of the two satisfying assignments of the verification equation (ii), we have:
\begin{align*}C = \sum_{k\in\{1,3,4\}}&{(c-c')^{-1}(s_k-s'_k) \cdot G_k} \\
                &+ (c-c')^{-1}(s_{5} - s'_{5}) \cdot H\end{align*}
The vector of coefficients above, where the coefficient of $G_2$ is taken to be zero, constitutes a valid witness for the relation. Observe that $(c-c')^{-1}(s_3 - s_3')$ is also a discrete logarithm of $\pk_{\prvr}$ to the base $G$. By taking the difference of the two satisfying assignments of verification equation (i), we see that:
\begin{alignat*}{2}
    &\quad &(c-c') \cdot \pk_{\prvr} &= (s_3-s'_3)\cdot G \\
    &\Rightarrow\quad &\pk_{\prvr} &= (c-c')^{-1}(s_3-s'_3)\cdot G
\end{alignat*} \qed

\subsubsection{HVZK} We construct a simulator, $\schemefont{Sim}$, that, on input $(C,\pk_{\prvr},\pparams)$, and challenge $c$, performs as follows:
\begin{enumerate}
    \item $\schemefont{Sim}$ samples $s_1,s_3,s_4,s_5 \sample \mathbb{Z}_q^{*}$ then computes:
    \begin{align*}
        t_1 &= s_3 \cdot G - c \cdot \pk_{\prvr}\\
        t_2 &= s_5 \cdot H + \sum_{k\in\{1,3,4\}}{s_k \cdot G_k} - c \cdot C
    \end{align*}
    \item Outputs transcript $(t_1,t_2,c,\{s_k\}_{k\in\{1,3,4,5\}})$.
\end{enumerate}
Clearly, the output of $\schemefont{Sim}$ constitutes a valid transcript.
To show the distribution of it is identical to that of real protocol executions, we show that there exists a bijection between the random coins of $\prvr$ and those of the simulator.
For a fixed instance challenge pair $((C, \pk_{\prvr},\pparams),c)$, observe that a real protocol execution is conditioned on the $\prvr$'s random coins $\alpha_1,\alpha_3,\alpha_4,\alpha_5$, whereas a simulated transcript is dependant on the simulator's random coins, $\alpha_1, \alpha_3,\alpha_4,\alpha_5$. Consider the bijection
\begin{multline*}
(\alpha_1,\alpha_3,\alpha_4,\alpha_5) \mapsto \\
(\alpha_1+m_1 \cdot c,\alpha_3 + m_3 \cdot c,\alpha_4 + m_4 \cdot c, \alpha_5 + r \cdot c)
\end{multline*}
 from the random coins of $\prvr$ to the random coins of the simulator.
When a simulator chooses the random coins $(\alpha_1+m_1 \cdot c,\alpha_3 + m_3 \cdot c,\alpha_4 + m_4 \cdot c, \alpha_5 + r \cdot c)$, it produces the same transcript of an honest $\prvr$ with random coins $(\alpha_1,\alpha_3,\alpha_4, \alpha_5)$.
Both of these occur with equal probability.
Therefore, over every possible transcript-challenge pair, the distributions are identical. \qed

\subsection{Security of $\pi_{open}(\mathbf{m})$}\label{proof-open}
The PoK $\pi_{open}(\mathbf{m})$ works as detailed in Section~\ref{pi-open}.
Below, we prove its properties.

\subsubsection{Completeness} Completeness holds since the verification equations are always satisfied for a valid witness, and hence the verifier accepts with probability 1.

\subsubsection{Special Soundness} Given two colliding transcripts for the same instance $(C,\pparams)$, we construct a valid witness for the given relation.
We denote the colliding transcripts as:
\[
(t_1,c,\{s_k\}_{k\in\{1,\dots,n\}}, s_x) \text{ and } (t,c',\{s'_k\}_{k\in\{1,\dots,n\}}, s'_{x}).
\]
By taking the difference of the two satisfying assignments of the verification equation (ii), we have:
\begin{align*}
C &= (c-c')^{-1} (s_x - s'_x) \cdot H\\
 &+ (c-c')^{-1} \sum_{k \in \{1 \dots n\}} (s_k - s_k') \cdot G_k
\end{align*}
By substituting with the responses $s_k$ and $s_x$, we get:
\begin{alignat*}{2}
&\quad &C &= (c - c')^{-1} (c - c') \cdot (r \cdot H + \sum_{k \in \{1 \dots n\}} m_k \cdot G_k) \\
&\Rightarrow\quad &C &= r \cdot H + \sum_{k \in \{1 \dots n\}} m_k \cdot G_k
\end{alignat*}\qed

\subsubsection{HVZK} We construct a simulator, $\schemefont{Sim}$, that on public input $(C, \pparams)$, and the challenge value $c$ performs as follows:

\begin{enumerate}
	\item $\schemefont{Sim}$ samples $(s_x, s_1, \dots s_k) \sample \mathbb{Z}_q^{*}$ and then computes:
		\begin{equation*}
        	  t_1 = s_x \cdot H +  \sum_{k\in n} s_k \cdot G_k - c \cdot C
    		\end{equation*}
	\item Outputs transcript $(t_1, c, s_x, \{s_k\}_{k\in n})$
\end{enumerate}

The transcript clearly satisfies the verification equation (i) and is a valid simulated transcript.
It remains to be shown that the distribution over the transcripts produced by the $\schemefont{Sim}$ are equal to the distributions produced by an honest prover and honest verifier.
We can show this via the same strategy as there is a bijection between every transcript produced by the honest prover and $\schemefont{Sim}$. $\prvr$'s random coins consist of $n+1$ values $\alpha_x, \alpha_1, \dots, \alpha_n$.
Given the fixed values, each tuple uniquely defines $t$, and the values $s_x, s_1, \dots s_n$.
Similarly, the simulator $\schemefont{Sim}$ outputs $s_x', \{s_k'\}_{k\in n}$.
Both random coins are chosen uniformly random over the same probability space.
Now consider the bijection
\begin{align*}
(s_x, \{s_k\}_{k\in n}) \mapsto (r \cdot c + \alpha_x, \{m_k \cdot c + \alpha_k\}_{\{k \in n\}}).
\end{align*}
When $\schemefont{Sim}$ chooses the random coins $(s_x, \{s_k\}_{k\in n})$, it produces the same transcript of an honest prover.
Both of these occur with equal probability.
Therefore, over every possible transcript-challenge pair, the distributions are equal.  \qed

\subsection{Security of $\pi_{add-mul}(m)$}\label{proof-add-mul}
The PoK $\pi_{add-mul}(m)$ works as detailed in Section~\ref{pi-add-mul}.
This proof can be reasoned as a composition between a PoK that proves an additive relationship (~\cref{pi-add}) and a PoK that proves a multiplicative relationship (~\cref{pi-mul}).
Let's explore this in detail and prove their properties.

\subsubsection{Proving that a value is the result of addition/subtraction}\label{pi-add}
This PoK aims to show that committed values ($C$) are the result of addition or subtraction operations.
These proofs (either $\pi_{add}(m)$ or $\pi_{sub}(m)$ for addition or subtraction, respectively) verify the following relation:

\scalebox{0.9}{%
\begin{minipage}{\columnwidth}
\begin{multline*}
\mathcal{R}_{\text{add/sub ($\pm$)}} =\\
\left\{
  \begin{array}{c}
    (C,\pparams),\\
    (\mathbf{x, y, (x \pm y), (r_1 \pm r_2)})\ \\
  \end{array}
\Bigg|
\begin{array}{c} \pparams = (G, H, q),\\
    (x, y, (x \pm y), (r_1 \pm r_2)), \\
C = (G \cdot x) + (H  \cdot {r_1}) \pm \\
(G \cdot y) + (H \cdot {r_2}) \\
\end{array}\right\}.
 \end{multline*}
\end{minipage}%
}

Given public parameters to the Pedersen commitment scheme $(G, H, q)$, which are points of an elliptic curve of order $q$, the interactive protocol works as follows.
Given public input $(C_1 = (G \cdot x) + (H \cdot {r_1}), C_2 = (G \cdot y) + (H \cdot {r_2}), C_3 = C_1 \pm C_2, \pparams)$, and private input $(x \pm y, r_1 \pm r_2)$:

\begin{enumerate}
    \item $\prvr$ samples $(\alpha_1, \ldots \alpha_6) \sample \mathbb{Z}_q^{*}$, then computes and sends:
        $(t_1 := \alpha_{1}\cdot G)$,
        $(t_2 := \alpha_{2}\cdot H)$,
        $(t_3 := G \cdot \alpha_3 + H \cdot \alpha_4)$,
        $(t_4 := G \cdot \alpha_5 + H \cdot \alpha_6)$.
    \item $\vrfr$ receives $(t_1, t_2, t_3, t_4)$, samples a challenge $c \sample \mathbb{Z}_q^{*}$, and sends it to $\prvr$.
    \item $\prvr$ computes:
        $(s_1 := c \cdot (x \pm y) + \alpha_1)$,
        $(s_2 := c \cdot (r_1 \pm r_2) + \alpha_2)$,
        $(s_3 := c \cdot x + \alpha_3)$,
        $(s_4 := c \cdot r_1 + \alpha_4)$,
        $(s_5 := c \cdot y + \alpha_5)$,
        $(s_6 := c \cdot r_2 + \alpha_6)$,
    $\prvr$ sends $(s_1, \ldots, s_6)$ to $\vrfr$.
    \item $\vrfr$ accepts iff the following equations hold:
    \begin{align}
        G \cdot {s_3} + H \cdot {s_4} &\stackrel{?}{=} (t_3) (C_1 \cdot c){}\\
        G \cdot {s_5} + H \cdot {s_6} &\stackrel{?}{=} (t_4) (C_2 \cdot c){}\\
        G \cdot {s_1} + H \cdot {s_2} &\stackrel{?}{=} (t_1 + t_2) + (C_3 \cdot c){}
    \end{align}
\end{enumerate}

\subsubsection{Completeness} Completeness holds since the verification equations are always satisfied for a valid witness, and hence the verifier accepts with probability 1.
This property holds as long as the same arithmetic operations (addition or subtraction) are consistently used across all equations.

\subsubsection{Special Soundness} Given two colliding transcripts for the same instance $(C,\pparams)$, we construct a valid witness for the
relation $\mathcal{R}_{\text{add/sub ($\pm$)}}$ above. We denote the colliding transcripts as:
\[
(t_1, \dots, t_4, c, s_1, \dots, s_6) \text{ and } (t_1, \dots, t_4, c',s_1', \dots, s_6').
\]
We define the output of an extractor to be $(x, r_1, y, r_2, x \pm y, r_1 \pm r_2)$:
\begin{align*}
x &= \frac{s_3-s_3'}{c-c'} & r_1 &= \frac{s_4-s_4'}{c-c'}& y &= \frac{s_5' - s_5}{c-c'}\\
r_2 &= \frac{s_6-s_6'}{c-c'}& x\pm y &= \frac{s_1-s_1'}{c-c'} & r_1\pm r_2 &= \frac{s_2-s_2'}{c-c'}
\end{align*}
We first show that the values $(x, r_1)$, $(y, r_2)$, are proper openings of the commitments to $C_1$ and $C_2$, respectively.
For the first, remember that:
\begin{align*}
G \cdot {s_3} + H \cdot {s_4} &\stackrel{?}{=} (t_3) (C_1 \cdot c){}\\
G \cdot {s'_3} + H \cdot {s'_4} &\stackrel{?}{=} (t_3) (C_1 \cdot c'){}
\end{align*}
It follows by dividing the two equations that:
\begin{align*}
C_1 &= G \cdot \{s_3 - s'_3\}/\{c-c'\} + H \cdot \{s_4 - s'_4\}/\{c-c'\}{}\\
C_1 &= G \cdot x + H \cdot r_1 \\
\end{align*}
which shows that $(x, r_1)$ is indeed an opening for commitment $C_1$.
For the second, remember that:
\begin{align*}
G \cdot {s_5} + H \cdot {s_6} &\stackrel{?}{=} (t_4) (C_2 \cdot c){} \\
G \cdot {s'_5} + H \cdot {s'_6} &\stackrel{?}{=} (t_4) (C_2 \cdot c'){}
\end{align*}
It follows by dividing the two equations that:
\begin{align*}
C_2 &= G \cdot \{s_5 - s'_5\}/\{c-c'\} + H \cdot \{s_6 - s'_6\}/\{c-c'\}{}\\
C_2 &= G \cdot y + H \cdot r_2 \\
\end{align*}
which shows that $(y, r_2)$ is indeed an opening for commitment $C_2$.
Finally for $C_3$, remember that:
\begin{align*}
G \cdot {s_1} + H \cdot {s_2} &\stackrel{?}{=} (t_1 + t_2) (C_3 \cdot c){}\\
G \cdot {s'_1} + H \cdot {s'_2} &\stackrel{?}{=} (t_1 + t_2) (C_3 \cdot c'){}\\
\end{align*}
It follows by dividing the two equations that:
\begin{align*}
C_3 &= G \cdot \{s_1 - s'_1\}/\{c-c'\} + H \cdot \{s_2 - s'_2\}/\{c-c'\}{}\\
C_3 &= G \cdot (x \pm y) + H \cdot (r_1 \pm r_2) \\
\end{align*}
which shows that $(x \pm y, r_1 \pm r_2)$ is indeed an opening for commitment $C_3$. \qed

\subsubsection{HVZK}
We construct a simulator, $\schemefont{Sim}$, that on public input $(C_1, C_2, C_3, \pparams)$, and the challenge value $c$ performs as follows:
\begin{enumerate}
	\item Samples $(s_1, \dots s_6) \sample \mathbb{Z}_q^{*}$ and then computes:
		\begin{align*}
			t_1 + t_2 &= s_1 \cdot G + s_2 \cdot H - (c \cdot C_3)\\
        		t_3 &= (s_3 \cdot G + s_4 \cdot H) \cdot (c \cdot C_1)^{-1}\\
			t_4 &= (s_5 \cdot G + s_6 \cdot H) \cdot (c \cdot C_2)^{-1}\\
    		\end{align*}
	\item Outputs transcript $(t_1 + t_2, t_3, t_4, c, s_1, \dots s_6)$
\end{enumerate}

The transcript clearly satisfies the verification equation (8-10) and is a valid transcript.
It remains to be shown that the distribution over the transcripts produced by $\schemefont{Sim}$ are equal to the distributions produced by an honest prover and honest verifier.
We can show this, by showing that there exists a bijection between every transcript produced by the honest prover and $\schemefont{Sim}$.
For a fixed relation $\mathcal{R}_{\text{add/sub ($\pm$)}}$, a witness, and a challenge c, the prover's random coins consist of $6$ values $\alpha_1, \dots, \alpha_6$.
Given the fixed values, each tuple uniquely defines $t_1, t_2, t_3, t_4$, and the values $s_1, \dots s_6$.
Similarly, $\schemefont{S}$ outputs $s_1', \dots, s_6'$.
Both random coins are chosen uniformly random over the same probability space.
Now consider the bijection:

$$
(s_1', \dots, s_6') \mapsto
\left(
\begin{aligned}
&c \cdot (x \pm y) + \alpha_1,\\
&c \cdot (r_1 \pm r_2) + \alpha_2,\\
&c \cdot x + \alpha_3,\\
&c \cdot r_1 + \alpha_4,\\
&c \cdot y + \alpha_5,\\
&c \cdot r_2 + \alpha_6
\end{aligned}
\right).
$$

When $\schemefont{Sim}$, chooses the random coins $(c \cdot (x \pm y) + \alpha_1, c \cdot (r_1 \pm r_2) + \alpha_2, c \cdot x + \alpha_3, c \cdot r_1 + \alpha_4, c \cdot y + \alpha_5, c \cdot r_2 + \alpha_6)$
, it produces the same transcript of an honest prover. Both of these occur with equal probability. Therefore,
over every possible transcript-challenge pair, the distributions are equal.\qed

\subsection{Proving that a value is the result of multiplication}\label{pi-mul}
This PoK aims to show that committed values ($C$) are the result of a multiplication operation.
We use the proof of [\cite{EPRINT:WTsTW17}, App. A] due to their efficiency, which is a folklore proof from~\cite{AFRICACRYPT:Maurer09}.
The proof ($\pi_{mul}(m)$) verify the following relation:

\[\mathcal{R}_{\text{mul}} = \left\{(C,\pparams),(x, y, (x * y))\ \Bigg|
\begin{array}{c} \pparams = (G, H, q),\\
    (x, y, (x \cdot y)), \\
C = (G \cdot (x \cdot y)) \\
\end{array}\right\}.\]

Given public parameters to the Pedersen commitment scheme $(G, H, q)$, which are points of an elliptic curve of order $q$, the interactive protocol works as follows.
Given public input $(C_1 = (G \cdot x) + (H \cdot {r_1}), C_2 = (G \cdot y) + (H \cdot {r_2}), C_3 = (G \cdot (x\cdot y)) + (H \cdot r_3), \pparams$, and private input $(x \cdot  y, x, y)$:

\begin{enumerate}
    \item $\prvr$ samples $(\alpha_1, \ldots \alpha_5) \sample \mathbb{Z}_q^{*}$, then computes and sends:
        $(t_1 := \alpha_1\cdot G + \alpha_2 \cdot H$),
        $(t_2 := \alpha_3\cdot G + \alpha_4 \cdot H$),
        $(t_3 := C_1 \cdot \alpha_3 + H \cdot \alpha_5$).
    \item $\vrfr$ receives $(t_1, t_2, t_3)$, samples a challenge $c \sample \mathbb{Z}_q^{*}$, and sends it to $\prvr$.
    \item $\prvr$ computes:
        $(s_1 := c \cdot x + \alpha_1)$,
        $(s_2 := c \cdot r_1 + \alpha_2)$,
        $(s_3 := c \cdot y + \alpha_3)$,
        $(s_4 := c \cdot r_2 + \alpha_4)$,
        $(s_5 := c \cdot (r_3 - r_1 \cdot y) + \alpha_5)$.
    $\prvr$ sends $(s_1, \ldots, s_5)$ to $\vrfr$.
    \item $\vrfr$ accepts if and only if the following equation holds:
    \begin{align}\tag{i}
        G \cdot {s_1} + H \cdot {s_2} &\stackrel{?}{=} (t_1) + (C_1 \cdot c){}\\
        G \cdot {s_3} + H \cdot {s_4} &\stackrel{?}{=} (t_2) + (C_2 \cdot c){}\\
        C_1 \cdot {s_3} + H \cdot {s_5} &\stackrel{?}{=} (t_3) + (C_3 \cdot c){}
    \end{align}
\end{enumerate}

The proof for Completeness, Special Soundness and HVZK follows from~\cite{EPRINT:WTsTW17}.\qed

\subsection{Security of $\pi_{add-mul(x)}$}\label{proof-add-mul}
As stated, the PoK $\pi_{add-mul}(x)$ works as a composition, and it is detailed in Section~\ref{pi-add-mul}.
The proof of Completeness, Special Soundness and HVZK follows from AND parallel composition of PoKs~\cite{dam-sigma}\qed.

\section{Decentralised KYC/Identity Proof}\label{appendix:identityproof}
Many financial institutions mandate \emph{Know Your Customer (KYC)} processes  for account management.
The purpose of these checks is to verify the users' identity, establish risk factors and prevent fraud.
Usually, identity verification is handled in a centralised way.
However, this diminishes users' control over their data and leads to privacy risks.
In the \protocolname protocol, users are given rewards and can verify their given rewards of the incentive system themselves.
To reduce the amount of fraud, and for users to verify their identities, the \protocolname protocol can require its users to additionally provide a proof of identity with each rewards request.

As KYC checks are cumbersome and due to legal requirements often require the presence of a human operator, we do not want to perform a KYC check for each reward request.
Hence, a KYC check can done once (and it is regularly re-done according to legal requirements), and then continuously identify a users wallet
by adding a zero-knowledge proof of a KYC'ed wallet with each following reward request.

The proposed KYC protocol works as follows:

\begin{enumerate}
    \item \IS creates an empty cryptographic accumulator and sends it to a layer 1 blockchain.
    \item The L1 blockchain stores the cryptographic accumulator in a smart contract.
    \item When using the system for the first time (or after a certain time has exceeded, due to some legal requirements), \U requests its wallet to be KYC'ed to the \IS or \AC.
    \item \IS or \AC requests to verify \U from the KYC provider.
    \item The KYC provider performs the KYC checks from \U asynchronously and according to the legal requirements necessary to be compliant.
    \item \U complies in the interactive KYC checks by providing its legal documents and other necessary details (such as a photograph for e.g. human verification).
    \item The KYC provider returns a report to \IS or \AC indicating if the KYC checks of \U have been successful or not.
    \item If the KYC checks have been successful, \IS or \AC updates the accumulator with the users' wallet ID.
    \item The L1 blockchain updates the accumulator in the smart contract with the added wallet ID.
    \item When performing a spending-reward request, \U additionally generates a zero-knowledge proof for their KYC'ed wallet.
    \item The L1 blockchain verifies the in the smart contract to check if \U's wallet is KYC'ed.
\end{enumerate}
\section{Security arguments}
\label{app:args}

We provide security arguments to protect the issuer $\IS$ of the incentive scheme from dishonest
users, as well as the other way around by arguing protections of honest users $\U$ from dishonest issuers.
Therefore, we show in the following how we can achieve our above mentioned goals to provide a secure
and privacy-preserving verifiable incentive scheme.

\subsubsection{Security of \protocolname}
Initially, we want to provide some security arguments to protect the issuer $\IS$ from dishonest users $\U$.
Therefore, we will show that the first three properties mentioned above hold. In more detail, we will
show that tokens are \emph{identity binding} with respect to the \emph{issuance procedure}, the balance
of tokens that a user can accumulate is \emph{balance binding} with respect to the \emph{collection} and
\emph{spending and public verifiability procedure}. Finally, we show that tokens can not be \emph{double-spend}
with respect to the \emph{spending and public verifiability procedure}.

\begin{algorithm}[!ht]
	\caption{Exp$_{\mathcal{A}, \text{issue}}^{\text{id-bind}}(\lambda)$}
	\KwOutput{The experiment returns 1 iff $\mathcal{A}$ made a successful call to MalIssue().}
	\hrule

	(sk$_{\IS}$, pk$_{\IS}) \leftarrow \IS.generate(1^{\$})$ \\
	(sk$_{\U}$, pk$_{\U}) \leftarrow \U.generate(1^{\$})$\\
	$b \leftarrow \mathcal{A}^{\text{MalIssue}}(pk_{\U})$
\end{algorithm}

\begin{algorithm}[!ht]
	\caption{Exp$_{\mathcal{A}, \text{collect, spend-verify}}^{\text{id-bind}}(\lambda)$}
	\KwInput{Token after the honest issuance procedure, or MalIssue has been called.}
	\KwOutput{The experiment returns 1 iff $\mathcal{A}$ made a successful call to MalCollect(v) or MalSpendVerify(v).}
	\hrule

	(sk$_{\IS}$, pk$_{\IS}) \leftarrow \IS.generate(1^{\$})$ \\
	(sk$_{\U}$, pk$_{\U}) \leftarrow \U.generate(1^{\$})$\\
	$b \leftarrow \mathcal{A}^{\text{MalCollect}}() || \mathcal{A}^{\text{MalSpendVerify}}()$
\end{algorithm}

\begin{definition}[Identity Binding~\cite{PoPETS:HKRR20}]
The \protocolname protocol is \emph{identity binding} if for any probabilistic polynomial time adversary
$\mathcal{A}$ in the experiment Exp$_{\mathcal{A}, \text{issue}}^{\text{id-bind}}(\lambda)$ and
Exp$_{\text{collect, spend-verify}, \mathcal{A}}^{\text{id-bind}}(\lambda)$ the advantages of the adversary defined by
$$
\text{Adv}_{\mathcal{A}, \text{issue}}^{\text{id-bind}}(\lambda) = \Pr \left[ \text{Exp}_{\mathcal{A}, \text{issue}}^{\text{id-bind}}(\lambda)  = 1 \right]
$$
\begin{multline*}
\text{Adv}_{\mathcal{A}, \text{collect, spend-verify}}^{\text{id-bind}}(\lambda) =\\ \Pr \left[ \text{Exp}_{\mathcal{A}, \text{collect, spend-verify}}^{\text{id-bind}}(\lambda)  = 1 \right]
\end{multline*}
are negligible in $\lambda$.
\end{definition}

\begin{algorithm}[!ht]
	\caption{Exp$_{\mathcal{A}, \text{\protocolname}}^{\text{bal-bind}}(\lambda)$}
	\KwOutput{The experiment returns 1 iff the following conditions are satisfied:
	\begin{itemize}
	\item all successful calls to MalIssue, MalCollect produced unique token id's and
	\item the claimed balance $w$ does not equal the sum of accumulated values $v$ for an particular user
	identified by it's public key $pk_{\U}$, i.e.,
	$$
	w \neq \sum_{v\in V_{\text{pk}_{\U}}} v,
	$$
	where $V_{\text{pk}_{\U}}$ is a collection of all previously processed values by successful calls to MalCollect for a unique $pk_{\U}$.
	\end{itemize}
	.}
	\hrule

	(sk$_{\IS}$, pk$_{\IS}) \leftarrow \IS.generate(1^{\$})$ \\
	$b \leftarrow \mathcal{A}^{\text{MalIssue, MalCollect, MalSpendVerify}}(pk_{\IS})$
\end{algorithm}

\begin{definition}[Balance Binding~\cite{PoPETS:HKRR20}]
The \protocolname protocol is \emph{balance binding} if for any probabilistic polynomial time adversary
$\mathcal{A}$ in the experiment Exp$_{\mathcal{A}, \text{\protocolname}}^{\text{bal-bind}}(\lambda)$ the
advantage of the adversary defined by
$$
\text{Adv}_{\mathcal{A}, \text{\protocolname}}^{\text{bal-bind}}(\lambda) = \Pr \left[ \text{Exp}_{\mathcal{A}, \text{\protocolname}}^{\text{bal-bind}}(\lambda)  = 1 \right]
$$
is negligible in $\lambda$.
\end{definition}

\begin{algorithm}[!ht]
	\caption{Exp$_{\mathcal{A}, \text{\protocolname}}^{\text{d-spend}}(\lambda)$}
	\KwInput{Tokens after two successful calls to MalCollect.}
	\KwOutput{The experiment returns 1 iff for two double spending tags \emph{dtag} and \emph{dtag}' at least one of the following conditions is satisfied:
	\begin{itemize}
		\item pk$_{\U} \neq$ pk$_{\U}$' or
		\item DedectDoubleSpend(pk$_{\IS}$, dtag, dtag') $\neq$ (pk$_{\U}, \pi_{open_{\text{sk}_{\U}}}$) or
		\item DedectDoubleSpend(pk$_{\IS}$, dtag, dtag') = (pk$_{\U}, \pi_{open_{\text{sk}_{\U}}}$) but $\pi_{open_{\text{sk}_{\U}}} = \bot$
	\end{itemize}
	.}
	\hrule
	(sk$_{\IS}$, pk$_{\IS}) \leftarrow \IS.generate(1^{\$})$ \\
	(sk$_{\U}$, pk$_{\U}) \leftarrow \U.generate(1^{\$})$\\
	$b \leftarrow \mathcal{A}^{\text{MalIssue}, \text{MalCollect}}(pk_{\IS})$
\end{algorithm}

\begin{definition}[Double-Spending Detection~\cite{PoPETS:HKRR20}]
The \protocolname protocol provides \emph{double-spending detection} if for any probabilistic polynomial
time adversary $\mathcal{A}$ in the experiment Exp$_{\mathcal{A}, \text{\protocolname}}^{\text{d-spend}}(\lambda)$
the advantage of the adversary defined by
$$
\text{Adv}_{\mathcal{A}, \text{\protocolname}}^{\text{d-spend}}(\lambda) = \Pr \left[ \text{Exp}_{\mathcal{A}, \text{\protocolname}}^{\text{d-spend}}(\lambda)  = 1 \right]
$$
is negligible in $\lambda$.
\end{definition}

\subsubsection{Privacy of \protocolname}
We also want to show that honest users $\U$ are protected against dishonest issuers $\IS$. Therefore,
we will show that the \emph{user privacy} is respected and the users operations are unlinkable over the
entire protocol flow. Additionally, we show that honest users are protected against \emph{false-accusations}
if they behave honest.

\begin{algorithm}[!ht]
	\caption{Exp$_{\mathcal{A}, \text{\protocolname}}^{\text{priv-ideal-world}}(\lambda)$}
	\hrule
	(sk$_{\IS}$, pk$_{\IS}) \leftarrow \mathcal{A}_{0}.generate(1^{\$})$ \\
	$b \leftarrow \mathcal{A}_{1}^{\text{SimIssue}, \text{SimCollect}, \text{SimSpendVerify}}(pk_{\IS})$
\end{algorithm}

\begin{algorithm}[!ht]
	\caption{Exp$_{\mathcal{A}, \text{\protocolname}}^{\text{priv-real-world}}(\lambda)$}
	\hrule
	(sk$_{\IS}$, pk$_{\IS}) \leftarrow \mathcal{A}_{0}.generate(1^{\$})$ \\
	$b \leftarrow \mathcal{A}_{1}^{\text{Issue}, \text{Collect}, \text{SpendVerify}}(pk_{\IS})$
\end{algorithm}

\begin{definition}[Privacy-Preserving~\cite{PoPETS:HKRR20}]
The \protocolname protocols are \emph{privacy-preserving} with respect to user \U privacy if there
exist probabilistic polynomial time algorithms SimIssue, SimCollect, SimSpendVerify
and for any probabilistic polynomial time adversary $\mathcal{A} = (\mathcal{A}_0, \mathcal{A}_1)$ in the experiment
Exp$_{\mathcal{A}, \text{\protocolname}}^{\text{privacy}}(\lambda)$
the advantage of the adversary defined by
\begin{align*}
\text{Adv}_{\mathcal{A}, \text{\protocolname}}^{\text{privacy}}(\lambda) &= \Pr \left[ \text{Exp}_{\mathcal{A}, \text{\protocolname}}^{\text{priv-real-world}}(\lambda)  = 1 \right] \\
&- \Pr \left[ \text{Exp}_{\mathcal{A}, \text{\protocolname}}^{\text{priv-ideal-world}}(\lambda)  = 1 \right]
\end{align*}
is negligible in $\lambda$.
\end{definition}

\begin{algorithm}[!ht]
	\caption{Exp$_{\mathcal{A}, \text{\protocolname}}^{\text{false-ac}}(\lambda)$}
	\KwInput{The double spending tag \emph{dtag} associated to the token, the proof $\pi_{open_{\text{sk}_{\U}}}$ from procedure DedectDoubleSpend.}
	\KwOutput{The experiment returns 1 iff the proof $\pi_{open_{\text{sk}_{\U}}}$ for pk$_{\U}$ verifies correctly indicating that the user is misbehaving.}
	\hrule
	(sk$_{\IS}$, pk$_{\IS}) \leftarrow \mathcal{A}_{0}.generate(1^{\$})$ \\
	(sk$_{\U}$, pk$_{\U}) \leftarrow \U.generate(1^{\$})$\\
	$b \leftarrow \mathcal{A}^{\text{Issue}, \text{Collect}}(pk_{\IS}, pk_{\U})$
\end{algorithm}

\begin{definition}[False-Accusation Protection~\cite{PoPETS:HKRR20}]
The \protocolname protocol provides \emph{false-acusation protection} if for any probabilistic polynomial
time adversary $\mathcal{A} = (\mathcal{A}_0, \mathcal{A}_1)$ in the experiment Exp$_{\mathcal{A}, \text{\protocolname}}^{\text{false-ac}}(\lambda)$
the advantage of the adversary defined by
$$
\text{Adv}_{\mathcal{A}, \text{\protocolname}}^{\text{false-ac}}(\lambda) = \Pr \left[ \text{Exp}_{\mathcal{A}, \text{\protocolname}}^{\text{false-ac}}(\lambda)  = 1 \right]
$$
is negligible in $\lambda$.
\end{definition}

\section{Additional Experiments}

 We provide the results of additional experiments in~\cref{table:bandwidth} and~\cref{table:indiv-bandwidth}.

\begin{table}[t]
  \ra{1.2}
\centering
  \resizebox{\columnwidth}{!}{%
  \begin{tabular}{c c r r r r r}
      \toprule
      {\(\curve\)} & {\(\textbf{M1}\)} & {\textbf{M2}} & {\textbf{M3}} & {\textbf{M4}} & {\textbf{M5}} &{\textbf{Populate-state}}\\
      \midrule
      \multicolumn{7}{c}{\textbf{Issuance}}\\
      \midrule
  \textbf{\secp} & 0.76413 & 1.0348 & 0.91592 & 2.1374e-5 & \text{-} & 0.45333 \\
  \textbf{\secq} & 0.7579 & 1.0420 & 0.92096 & 2.0547e-5 & \text{-} & 0.44408\\
  \midrule
  \multicolumn{7}{c}{\textbf{Collection}}\\
  \midrule
  \textbf{\secp} & 0.00076972 & 3.2768 & 4.2139 & 0.93403 & 2.1655e-5 & 0.4513 \\
  \textbf{\secq} & 0.00072432 & 3.3803 & 4.6076 & 0.91914 & 2.1656e-5 & 0.46515 \\
  \midrule
  \multicolumn{7}{c}{\textbf{Spending}}\\
  \midrule
  \textbf{\secp} & 0.00076881 & 34.093 & 20.144 & 1.6359 & 2.1154e-5 & 0.45203\\
  \textbf{\secq} & 0.00072669 & 34.606 & 21.001 & 1.71452 & 2.1091e-5 & 0.45073\\
  \bottomrule
  \multicolumn{7}{c}{\textbf{Spending-Verify}}\\
  \midrule
  \textbf{\secp} & 0.00076881 & 34.093 & 40.354 & 3.6289 & 2.1154e-5 & 0.45203\\
  \textbf{\secq} & 0.00072669 & 34.606 & 40.555 & 3.7135 & 2.1091e-5 & 0.45073\\
  \bottomrule
  \end{tabular}
  }
\caption{Performance ($\SI{}{\milli\second}$) on Apple M3, 24$\SI{}{\giga\byte}$ of memory, for \protocolname.}
\label{table:bandwidth}
\end{table}

\begin{table}[t]
  \ra{1.2}
  \centering
  \resizebox{\columnwidth}{!}{%
  \begin{tabular}{c c r r r r}
    \toprule
    {\(\textbf{Primitive}\)} &\multicolumn{5}{c}{\textbf{Operations}}\\
    \midrule
    {\textbf{ALC}} & $\mathsf{Comm}$ & $\mathsf{Chall}$ & $\mathsf{Resp}$ & $\mathsf{Sign}$ & $\mathsf{Verify}$ \\
    \midrule
    \textbf{\secp} & 980.31 & 1.9561 & 4.2369e-5 & 1.2668 & 1.2912 \\
    \textbf{\secq} & 1.0691 & 2.1016 & 3.8432e-5 & 1.3779 & 1.3828 \\
    \midrule
  \end{tabular}
  }

  \resizebox{\columnwidth}{!}{%
  \begin{tabular}{p{2cm} p{2cm} p{2cm}}
    \midrule
    {\textbf{ALC Proof}} & $\mathsf{Prove}$ & $\mathsf{Verify}$ \\
    \midrule
    \textbf{\secp} & 1.4615 & 1.0924 \\
    \textbf{\secq} & 1.5948 & 1.1924 \\
    \midrule
    {\textbf{Opening Proof}} & $\mathsf{Prove}$ & $\mathsf{Verify}$ \\
    \midrule
    \textbf{\secp} & 0.65882 & 0.79591 \\
    \textbf{\secq} & 0.35327 & 0.49711 \\
    \midrule
    {\textbf{Issuance Proof}} & $\mathsf{Prove}$ & $\mathsf{Verify}$ \\
    \midrule
    \textbf{\secp} & 0.82411 & 0.001118 \\
    \textbf{\secq} & 0.89891 & 1.2175 \\
    \midrule
    {\textbf{Add-Mul Proof}} & $\mathsf{Prove}$ & $\mathsf{Verify}$ \\
    \midrule
    \textbf{\secp} & 1.6436 & 2.3650 \\
    \textbf{\secq} & 1.7668 & 2.5448 \\
    \midrule
    {\textbf{Sub Proof}} & $\mathsf{Prove}$ & $\mathsf{Verify}$ \\
    \midrule
    \textbf{\secp} & 64.395 & 5.1636 \\
    \textbf{\secq} & 69.783 & 5.6046 \\
    \midrule
    {\textbf{Rewards Proof}} & $\mathsf{Prove}$ & $\mathsf{Verify}$ \\
    \midrule
    \textbf{\secp} & 69.665 & 5.8863 \\
    \textbf{\secq} & 75.342 & 6.3754 \\
    \midrule
    {\textbf{Membership Proof}} & $\mathsf{Prove}$ & $\mathsf{Verify}$ \\
    \midrule
    \textbf{\secp} & 830.53 & 18.459 \\
    \textbf{\secq} & 830.53 & 18.459 \\
    \bottomrule
  \end{tabular}
  }

  \caption{Performance ($\SI{}{\milli\second}$) of individual primitives on ``t2.2xlarge'' machine for \protocolname. Note that the ``membership proof'' is over $2^{20}$ elements.}
  \label{table:indiv-bandwidth}
\end{table}

\section{Public Verification via L1 Smart Contracts}\label{appendix:smartcontracts}

Some of the intended usecases for the \protocolname protocol are targeting large scale
systems with potentially millions of users. Therefore, the system scalability is of
critical importance. While the interaction with the incentive scheme to earn rewards
does not require a lot of resources and can easily be managed by a backend system,
the public verification of the rewards request on a single-threaded backend can
become a bottleneck.

For transparency and public verifiability the reward verification can be implemented in a smart contract on a L1 blockchain.
The choice of the L1 blockchain is an important decision, as for example Ethereum-based blockchains using the Ethereum Virtual Machine (EVM)~\cite{eth-vm} runtime are single-threaded.
However, the Solana blockchain~\cite{solana} uses a parallel smart contract run-time called Sealevel, that is only limited by the number of cores available in a validator~\cite{Yakovenko2019}.

Currently, Solana has no support for some of the underlying cryptography needed.
Furthermore, Solana currently only supports curve25519-ristretto for elliptic curves, while \protocolname needs a 2-cycle elliptic curve.
Yet, for our experimental analysis, we still implemented a native verification of only $\pi_{reward}$ in a Solana smart contract, and provide some numbers for system performance, scalability and financial costs.

\subsection{System Performance}

\begin{table}[!ht]
  \centering
  \resizebox{\columnwidth}{!}{%
    \begin{tabular}{c|cc}
        \toprule
        \textbf{Catalogue size} & \textbf{$\pi_{\text{reward}}$ proof} & \textbf{$\pi_{\text{reward}}$ verif} \\
        \midrule
        64  & 5.41 & 1.30 \\
        128 & 8.94 & 2.04 \\
        256 & 15.82 & 3.42 \\
        \bottomrule
    \end{tabular}%
  }
  \caption{Execution times ($\SI{}{\milli\second}$) for generating a single $\pi_{\text{reward}}$ and verification of it for different sizes of the incentive catalogue.}
  \label{tab:rewards-request}
\end{table}

Initially, we are interested in the time it takes to generate a reward request.
Table~\ref{tab:rewards-request} shows $\pi_{reward}$'s generation (remember that this proof executes a range and linear proof inside) and verification times for one single execution with various catalogue sizes (64, 128 and 256 incentives).
As we can see from the table, the generation for an incentive catalogue of 256 incentives takes approximately 16 $\SI{}{\milli\second}$.
We were also interested in its sizes so we can better understand the bandwidth requirements to process them.
Table~\ref{tab:proofsizes} shows $\pi_{reward}$ sizes for different sizes of the incentive catalogue.

\begin{table}[t]
  \centering
  \resizebox{\columnwidth}{!}{%
    \begin{tabular}{c|ccc}
        \toprule
        \textbf{Catalogue size} & \multicolumn{3}{c}{\textbf{$\pi_{\text{reward}}$}} \\
        ~ & range-proof & linear-proof & total \\
        \midrule
        64 & 480 & 480 & 960 \\
        128 & 480 & 544 & 1024 \\
        256 & 480 & 608 & 1088 \\
        \bottomrule
    \end{tabular}%
  }
  \caption{Rewards proof sizes ($\SI{}{\byte}$) for various incentive catalogue sizes. Note that the range-proofs are a constant size, as the limit of the maximum rewards does not change.}
  \label{tab:proofsizes}
\end{table}

\begin{figure}[t]
\centering
\includegraphics[width=0.5\textwidth]{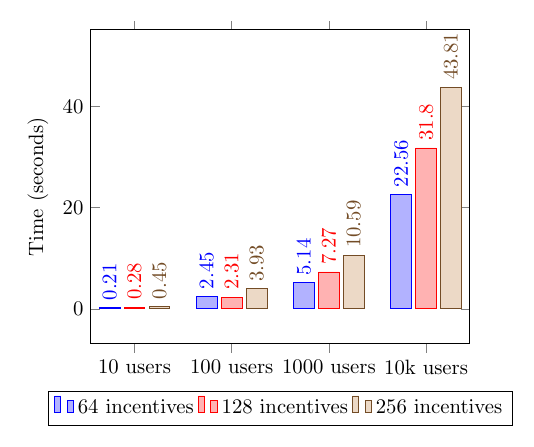}
\caption{Execution time of end-to-end reward requests (proof generation \&
verification) for multiple users and different sizes of the incentive catalogue. }
\label{fig:end2end}
\end{figure}

Finally, we explored the end-to-end performance of $\pi_{reward}$'s generation and verification by running in parallel multiple clients and using a single-threaded server.
Figure~\ref{fig:end2end} gives some performance numbers of the execution time using multiple users and different sizes of the incentives catalogue.

\subsection{System Scalability}

\begin{table}[t]
  \centering
  \resizebox{\columnwidth}{!}{%
    \begin{tabular}{cccc}
        \toprule
        \textbf{\# users} &
        \textbf{Size of data} &
        \multicolumn{2}{c}{\textbf{$\pi_{reward}$'s verification}} \\
        ~ & ~ & \textbf{t2.large} & \textbf{Solana}\\ \midrule
        1000 & $\SI{1.08}{\mega\byte}$ & $\SI{3.65}{\second}$ & $\SI{5.57}{\second}$\\
        $\num{10e3}$ & $\SI{10.88}{\mega\byte}$ & $\SI{36.47}{\second}$ & $\SI{55.73}{\second}$\\
        $\num{200e3}$ & $\SI{217}{\mega\byte}$ & $\SI{12}{\minute} \SI{9}{\second}$ & $\SI{18}{\minute} \SI{35}{\second}$\\
        $\num{5e6}$ & $\SI{5.44}{\giga\byte}$ & $\SI{5}{\hour} \SI{3}{\minute} \SI{54}{\second}$ & $\SI{7}{\hour} \SI{44}{\minute} \SI{25}{\second}$\\
        \midrule
        $\num{15.5e6}$ & $\SI{16.87}{\giga\byte}$ & \text{-} & $\SI{24}{\hour}$\\
        $\num{23.6e6}$ & $\SI{25.8}{\giga\byte}$ & $\SI{24}{\hour}$ & \text{-} \\
         \bottomrule
    \end{tabular}
  }
  \caption{Verification of rewards proofs for multiple users on a ``t2.large'' server and a Solana validator as native programs. The incentive catalogue is fixed to 256 entries.}
   \label{tab:proofverifcomp}
\end{table}

We implemented $\pi_{reward}$'s verification and run benchmarks in a single-threaded server environment.
In our experiments, we fixed the incentive catalogue to 256 entries, which we deem fits the most practical usecases.
Table~\ref{tab:proofverifcomp} shows the results, indicating that using our not optimised implementation we can verify up to 23.6 million reward requests in one day.
However, an interesting observation and practical bottleneck is that the server would need to process roughly 25.8 GB of data from the proof sizes alone.

Initially, we implemented the rewards proof verification as an on-chain program,
running it inside the Solana VM, and doing syscalls for the elliptic curve operations
outside on the validator. Unfortunately, currently there is no support for the
extendable-output function (XOF) hash operations needed in the
merlin\footnote{\url{https://github.com/dalek-cryptography/merlin}} transcripts that
implement the Fiat-Shamir transform (making an interactive zero-knowledge proof
non-interactive). Consequently, we implemented the rewards proof verification
as a native program, running entirely on the validator. Table~\ref{tab:proofverifcomp}
shows the results, indicating that on the Solana blockchain, 15.5 million reward
requests can be verified daily.


\subsection{Financial Costs}

\begin{table}[htb]
  \centering
  \resizebox{\columnwidth}{!}{%
    \begin{tabular}{lccc}
        \toprule
        ~ & \multicolumn{3}{c}{\textbf{Financial costs (US\$)}} \\
        \textbf{Machine} & \textbf{Computation} & \textbf{Data processing} & \textbf{Total} \\ \midrule
        \textbf{t2.large} & 0.47 & 0 & 0.47\\
        \textbf{Solana} & 3545 & 0 & 3545\\
        \bottomrule
    \end{tabular}
  }
  \caption{Financial cost comparison of running rewards verification on a server versus on-chain in a smart contract of a L1 blockchain. The estimates are based on $\num{5e6}$ users, processing $\SI{5.44}{\giga\byte}$ data in and $\SI{195}{\mega\byte}$ data out.}
   \label{tab:costs}
\end{table}

We calculate the server-side costs for running the rewards proof verification, for two possible deployment scenarios: i) when the rewards
verification is done on a ``t2.large'' server, and ii) when the rewards verification is done on-chain on a L1 blockchain within a smart contract. Table~\ref{tab:costs} provides estimates for both scenarios.

Using a AWS EC2 ``t2.large'' server with on-demand pricing, the computation time for verifying the reward requests of $\num{5e6}$ users is $\SI{5.065}{\hour}$, with a hourly rate of 0.0928 US\$, so the costs are 0.47 US\$.
AWS also charges for processing data, however, processing incoming data is free, while outgoing data costs approximately 0.09 US\$ per $\SI{}{\giga\byte}$ (with the first $\SI{100}{\giga\byte}$ free per month).
The response length of one reward request is only $\SI{39}{\byte}$, leading to $\SI{195}{\mega\byte}$ for $\num{5e6}$ users, which is within the free range.

Using the Solana blockchain to run the rewards proof verification in a smart contract we can execute the whole smart contract in one transaction, as the transaction limit is $\SI{1.2}{\kilo\byte}$ per transaction.
We do not need to store any data in an account, therefore we can avoid any storage rent costs.
For an incentive catalogue size of 256 entries, the proof sizes are $\SI{1.08}{\kilo\byte}$ which fits in one transaction. The total proof
verification, would need to execute one transaction and cost 5000 lamports (equivalent to 0,000709 US\$\footnote{Exchange rate as of July 10th 2024, 12:00 UTC}).
The costs to verify proofs of $\num{5e6}$ users, would be 3545 US\$ for computation, and no costs for processing the data.



\end{document}